\documentclass[twocolumn, trackchanges]{aastex62}

\usepackage{url}
\usepackage{amsmath}
\usepackage{amssymb}
\usepackage{natbib}
\usepackage{multirow}
\usepackage{xcolor}
\usepackage{lineno}

\newcommand{\bvec}[1]{{\ensuremath{\boldsymbol{#1}}}}

\graphicspath{}

\received{}
\revised{}
\accepted{}
\submitjournal{}

\shorttitle{Mutual Transits}
\shortauthors{Gordon et al.}

\begin{document}

\title{Analytic Light Curve for Mutual Transits of Two Bodies Across a Limb-darkened Star}

\correspondingauthor{Tyler A. Gordon}
\email{tagordon@uw.edu}

\author[0000-0001-5253-1987]{Tyler A. Gordon}
\affiliation{Department of Astronomy, University of Washington, Box 351580, U.W., Seattle, WA 98195-1580, USA}

\author[0000-0002-0802-9145]{Eric Agol}
\affiliation{Department of Astronomy and Virtual Planetary Laboratory, University of Washington, Box 351580, U.W., Seattle, WA 98195-1580, USA}



\begin{abstract}
We present a solution for the light curve of two bodies mutually transiting a star with polynomial limb darkening. The term ``mutual transit'' in this work refers to a transit of the star during which overlap occurs between the two transiting bodies. These could be an exoplanet with an exomoon companion, two exoplanets, an eclipsing binary and a planet, or two stars eclipsing a third in a triple star system. We include analytic derivatives of the light curve with respect to the positions and radii of both bodies. We provide code that implements a photodynamical model for a mutual transit. We include two dynamical models, one for hierarchical systems in which a secondary body orbits a larger primary (e.g.\ an exomoon system) and a second for confocal systems in which two bodies independently orbit a central mass (e.g.\ two planets in widely separated orbits). Our code is fast enough to enable inference with MCMC algorithms, and the inclusion of derivatives allows for the use of gradient-based inference methods such as Hamiltonian Monte Carlo. While applicable to a variety of systems, this work was undertaken primarily with exomoons in mind. It is our hope that making this code publicly available will reduce barriers for the community to assess the detectability of exomoons, conduct searches for exomoons, and attempt to validate existing exomoon candidates. We also anticipate that our code will be useful for studies of planet-planet transits in exoplanetary systems, transits of circumbinary planets, and eclipses in triple-star systems.
    
\end{abstract}


\section{introduction}\label{sec:intro}

    Mutual transits, which have also been referred to as ``planet-planet eclipses'', ``planet-planet transits'', and ``overlapping double transits'' \citep{Luger2017, Hirano2012, Masuda2013, Masuda2014}, occur when two bodies transit a star at the same time and exhibit some degree of overlap during the simultaneous transit. This can happen between an exoplanet and its moon, between two exoplanets, between a planet and a star in a circumbinary system, or between two stars in an eclipsing triple-star system. In past work, different techniques have been used to model the light curves resulting from this three-body overlap depending on the type of system being investigated.
    
\begin{figure*}
                \centering
                \includegraphics[width=\hsize]{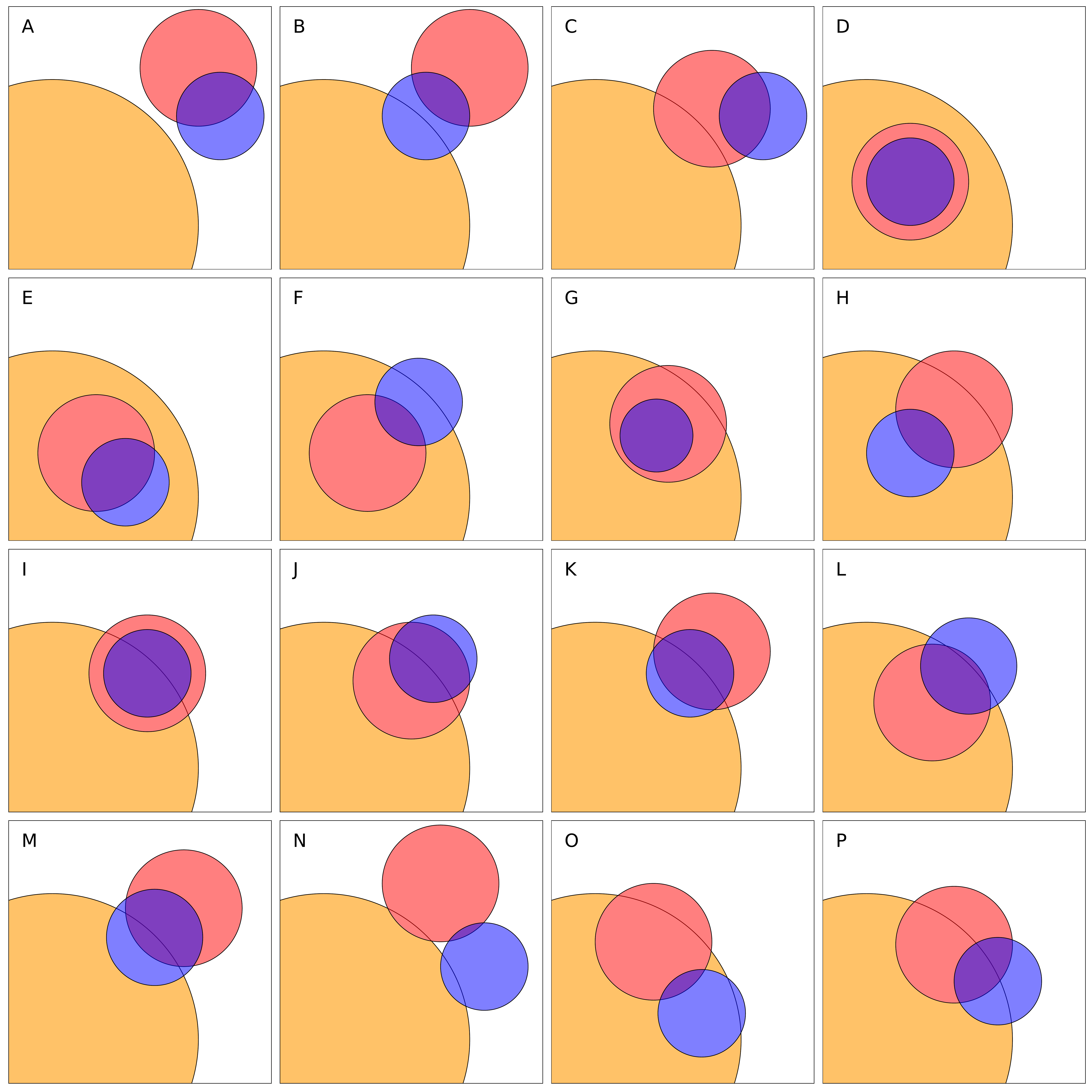}
                \caption{Representative examples of the different possible configurations of the planet-moon transit for the case where the planet and moon overlap each other. The letters in the top right of each box refer to the outcomes of the decision tree in Figure \ref{fig:flowchart}.} 
                \label{fig:configurations}
            \end{figure*}

    In this paper we present a solution for the flux blocked by two mutually transiting bodies. Our solution follows the method outlined by \citet{pal2012} in applying Green's theorem to transform the integral over the area of the star blocked by the transiting bodies into a line integral along the boundary of the unocculted area of the star, but unlike \citet{pal2012} which gives a general solution for N overlapping bodies, our solution is for exactly two bodies. This allows us to write down an analytic solution for the flux in each of the 16 possible geometric cases shown in Figure \ref{fig:configurations}. We also expand on previous work by computing derivatives with respect to the limb-darkening parameters and the positions and radii of each body.
    
    We use our solution to implement a photodynamical model which takes as input the dynamical parameters of two bodies, computes the positions of each at an array of times, and then calculates the flux observed at those times. We call this model \texttt{gefera}\footnote{\textit{gefera} is an Old English word with the literal meaning ``fellow traveller,'' which might be translated today as companion or comrade. We feel that this is an appropriate name for a code developed to search for exomoons, companions and fellow travellers to their exoplanet hosts.}. \texttt{Gefera} includes two dynamical models: a hierarchical model that can be used to simulate either an exomoon system or a hierarchical triple-star system, and a confocal model which is applicable to mutual transits of two exoplanets. In Figure \ref{fig:demo} we present a sample light curve for a mutual exomoon/exoplanet transit computed with \texttt{gefera}. In addition to computing the light curve itself, \texttt{gefera} provides derivatives of the light curve with respect to all of the input parameters. These derivatives can help speed up fitting algorithms and is essential for gradient-based MCMC methods such as Hamiltonian MCMC. Derivatives are also necessary to compute the information matrix, which gives insight into the information content of the light curve with respect to each parameter under given noise assumptions \citep{Vallisneri2008}. Figure \ref{fig:derivatives} shows the derivatives of the simulated light curve in Figure \ref{fig:demo}. \texttt{Gefera} has been developed openly in a public github repository and is provided for use by the community under a GPL license \footnote{The code is available at https://github.com/tagordon/gefera and is pip-installable under the name gefera. The Gnu Public License (GPL) generally permits both non-commericial and commerial use of the code under the stipulation that any works making use of the code are licensed in such a way that they grant users the same freedoms as received under the Gnu Public License. A permanent snapshot of the code is available on Zenodo \citep{gefera_zenodo}}. 
    
    In the remainder of the introduction we discuss existing models for mutual transits and the types of systems to which they have been applied, highlighting the differences between these models and \texttt{gefera}. 
    
    \subsection{Exomoon Transits}
    
    Despite a lack of observed exomoon transits, it is likely that these are the most frequently occurring, though not the most easily observable, mutual transit events. Several candidate exomoons have been identified in Kepler photometry (see \citealt{Teachey2018, Kipping2022}). The analyses that produced these candidates used the code \texttt{LUNA} \citep{Kipping2011} which employs the small-planet approximation (or in this case the small-moon approximation) from \cite{Mandel2002} to compute the flux blocked by the moon while applying the exact analytic solution (also from \citealt{Mandel2002}) for the planet.
    
    In contrast to \texttt{LUNA}, our method is exact for both bodies. While the small-planet approximation is very accurate for reasonably-sized exomoons, this does limit the utility of \texttt{LUNA} for very large satellites or binary planet pairs, especially around M-dwarfs with small radii. \texttt{LUNA} does not compute derivatives of the transit light curve. Finally, \texttt{LUNA} is proprietary whereas \texttt{gefera} is being made publicly available.
    
    After this paper was initially submitted the pubicly available exomoon transit code \texttt{Pandora} \citep{Hippke2022} was released. Like \texttt{LUNA}, this code implements an approximate solution to the mutual transit based on the small-planet approximation. However unlike \texttt{LUNA} the small-planet approximation is only applied to the area of overlap between the planet and the moon, so that the flux is exact when there is no overlap between the two bodies. \texttt{Pandora} does not compute derivatives of the light curve.
    
    \subsection{Mutual Transits of Exoplanets}
    
    A far less likely but more easily observed scenario is that in which two exoplanets overlap during a simultaneous transit \citep{Ragozzine2010}. \citet{Brakensiek2016} gives a full treatment of the geometric probability of these events. Such an event is thought to have been observed in the Kepler-94 system \citep{Hirano2012, Masuda2013}. Additionally, one observed simultaneous transit of Kepler-51 b and d shows a feature which may be explained by planet-planet overlap during the transit \citep[][to be further explored in \S\ref{sec:kepler51}]{Masuda2014}. 
    These authors model the mutual transit event as a superposition of two single transits computed using the \cite{Mandel2002} solution, with the addition of a ``bump'' function that depends on the area of the overlap between the two bodies and takes into account stellar limb-darkening in an approximate fashion similar to the small-planet approximation put forth in \cite{Mandel2002}. The difference between the methods used by \citet{Masuda2013, Masuda2014} and \cite{Kipping2011} is that in the latter method the small-planet approximation for the entire area of the smaller body, whereas \citet{Masuda2013} and \citet{Masuda2014} only approximate the flux blocked by the region of overlap between the two bodies -- the non-overlapping regions are computed exactly.
    
    The code \texttt{planetplanet} described in \cite{Luger2017} is also capable of modeling planet-planet mutual transits. This code implements a somewhat novel integration scheme in which the limb-darkened stellar surface is approximated by a series of annuli of constant radiance. The region of overlap between these annuli and the occulting body can then be computed analytically. This method is then generalized for overlapping bodies. This treatment of the overlapping transit is analytic in the sense that the integral for each annulus is computed analytically, but relies on an approximation of the limb-darkening profile and requires a potentially large number of evaluations of the analytic integrals to accurately approximate the limb-darkened light curve.
    
    \texttt{Gefera} again distinguishes itself in that it computes the exact light curve and includes derivatives, although in the case of the \citet{Masuda2013} method the light curve will be exact for any times during which the two bodies are non-overlapping. The derivatives enable faster optimization and inference than can be accomplished otherwise.
    
    \subsection{Eclipsing Triple-star Systems}
    
    The type of mutual transit event for which we have the most examples are those that take place in triple-star systems when two stars transit the primary. These eclipses have most frequently been modeled using the Wilson-Devinney \citep{Wilson1971} code or more recent codes based on that work (see \citealt{Borkovits2020, Borkovits2022} and \citealt{Mitnyan2020} for some recent examples that make use of \texttt{lightcurvefactory}, \citealt{Borkovits2013, Borkovits2019a}). These codes compute light curves numerically, accounting for tidal distortion and light travel time, among other effects, and often including radial velocities and spectral information in their output. While slower to compute than an analytical model, the inclusion of tidal distortion and light travel time effects is important for accurately modeling many multi-star systems.  
    
    When spectral data is not available and when tidal distortion effects are not considered important, much faster analytic methods can be employed. \cite{Carter2012} use an analytic method developed by \cite{pal2012} which employs Green's theorem to transform the integral over the occulted region of the star into a line integral along the boundary of the occulter. The code \texttt{photodynam}, based on that method, has been used to model the Kepler-126 system, a hierarchical triple-star system in which two low mass stars transit a third star \citep{Carter2011}. \texttt{Photodynam} has also been used in the analysis of circumbinary planets Kepler-34 b and Kepler-35 b \citep{Welsh2012}, Kepler-16 b \citep{Doyle2011} and Kepler-36 b and c \citep{Carter2012}. \cite{Short2018} also model Kepler-126 and Kepler-16 by numerically evaluating the line integral given in \cite{pal2012}. This allows for the use of arbitrary limb-darkening laws and enables them to model the Rossiter-McLaughlin effect using the same principle of numerically evaluating the line integral over a map of the radial velocity.
    
    Although \texttt{gefera} is not a viable alternative to the full spectral model implemented by \citet{Wilson1971} and similar codes, it may be preferable to \texttt{photodynam} in cases where the more complete model is not required. While both \texttt{photodynam} and \texttt{gefera} compute an exact light curve, \texttt{photodynam} does not include derivatives and was found to be slightly slower than \texttt{gefera} in testing. Finally, whereas \texttt{photodynam} is available only as a \texttt{c/c++} code and is coded to write output to a file, \texttt{gefera} is designed from the ground up for Bayesian inference and includes an easy-to-use \texttt{python} interface. \texttt{Gefera} allows for one or both of the transiting bodies to be larger than the star. For transiting triple-star systems the user may wish to add the flux from the transit between the two foreground stars, which can be accomplished by running \texttt{gefera} a second time with the radius of one of the transiting bodies set to zero and adding this to the flux from the three-body computation. Details on this procedure are given in section \ref{sec:triplestars}
    
    \begin{figure*}
        \centering
        \includegraphics[width=\textwidth]{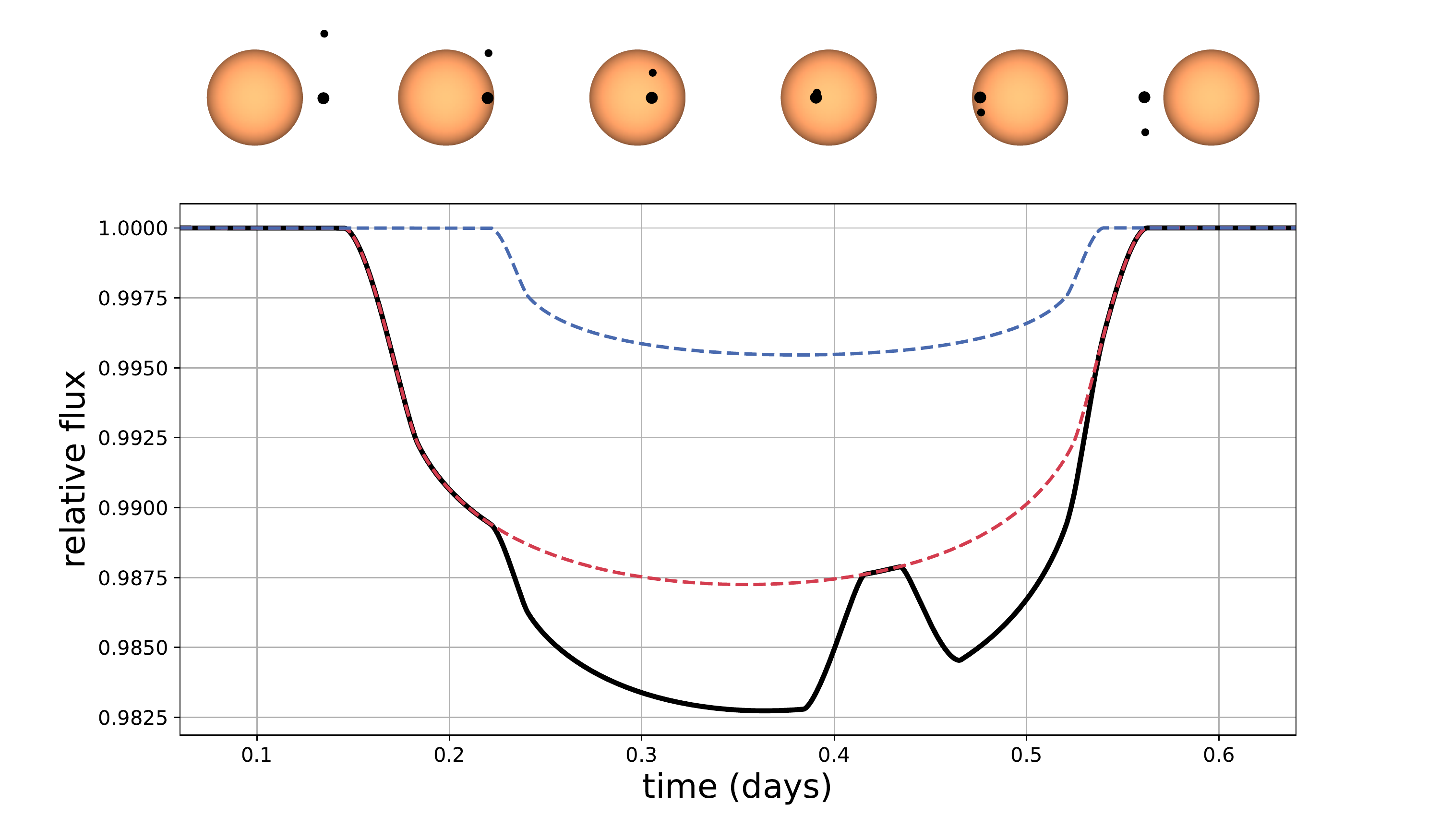}
        \caption{A simulated exoplanet/exomoon transit in which the exomoon and exoplanet fully overlap while transiting the star. The snapshots across the top of the plot show the configuration of the system at the points in time indicated along the x-axis of the light curve plot. The blue dashed line shows the transit of the moon as it would appear if it were the only transiting body, and the red dashed line shows the same for the planet.}
        \label{fig:demo}
    \end{figure*}
    
    \begin{figure*}
        \centering
        \includegraphics[width=\textwidth]{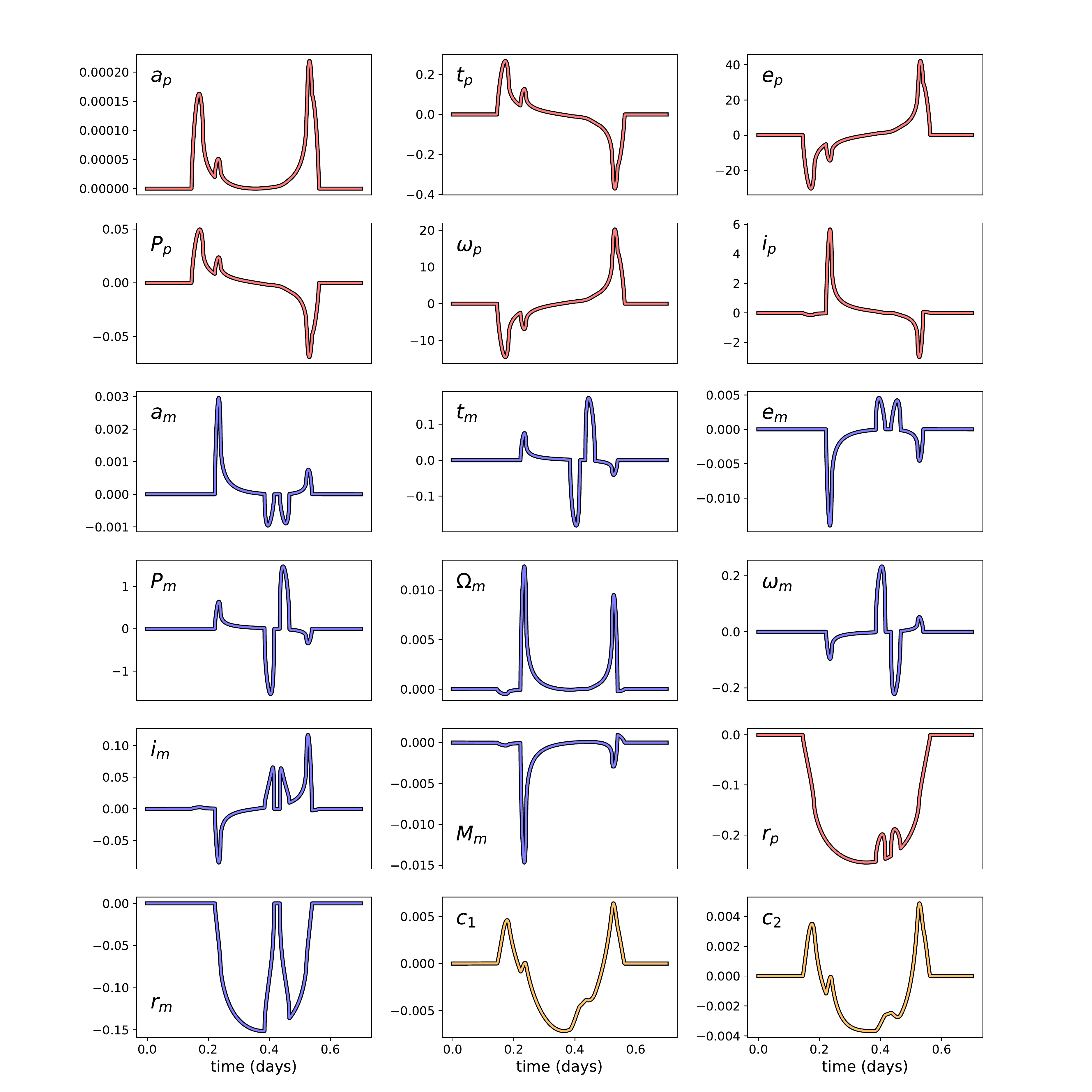}
        \caption{Derivatives of the simulated transit in Figure \ref{fig:demo} with respect to all model parameters.
        In order from left to right, top to bottom the parameters are the semimajor axis, epoch, eccentricity, period, longitude of periastron, and inclination of the planet, the semimajor axis, epoch, eccentricity, period, longitude of the ascending node, longitude of periastron, and inclination for the moon, the moon mass as a fraction of the planetary mass, the planetary radius, the moon's radius, and the two limb-darkening parameters.
        The units along the y-axes are in units of flux per the relevant unit for the parameter in question. For example, if the semimajor axis of the planet is in terms of stellar radii then the units for the upper left plot are flux per unit stellar radius. 
        The lines are color-coded as follows: parameters related to the planet are shown in red, those related to the moon are shown in blue, and the limb-darkening parameters for the star are shown in yellow.}
        \label{fig:derivatives}
    \end{figure*}
    
    As TESS and CHEOPS continue to collect light curves the small number of mutually transiting multi-planet systems is likely to expand. Furthermore, upcoming transit missions such as PLATO and ARIEL represent new opportunities to detect transiting exomoons, binary planets, or other interesting and exotic systems. These prospects all point towards the need for a fast, accurate, robust, and easy-to-use mutual transit model that is compatible with cutting-edge inference techniques. It is this need that we hope \texttt{gefera} will fulfill.
    
\begin{figure*}
                \centering
                \includegraphics[height=0.9\vsize]{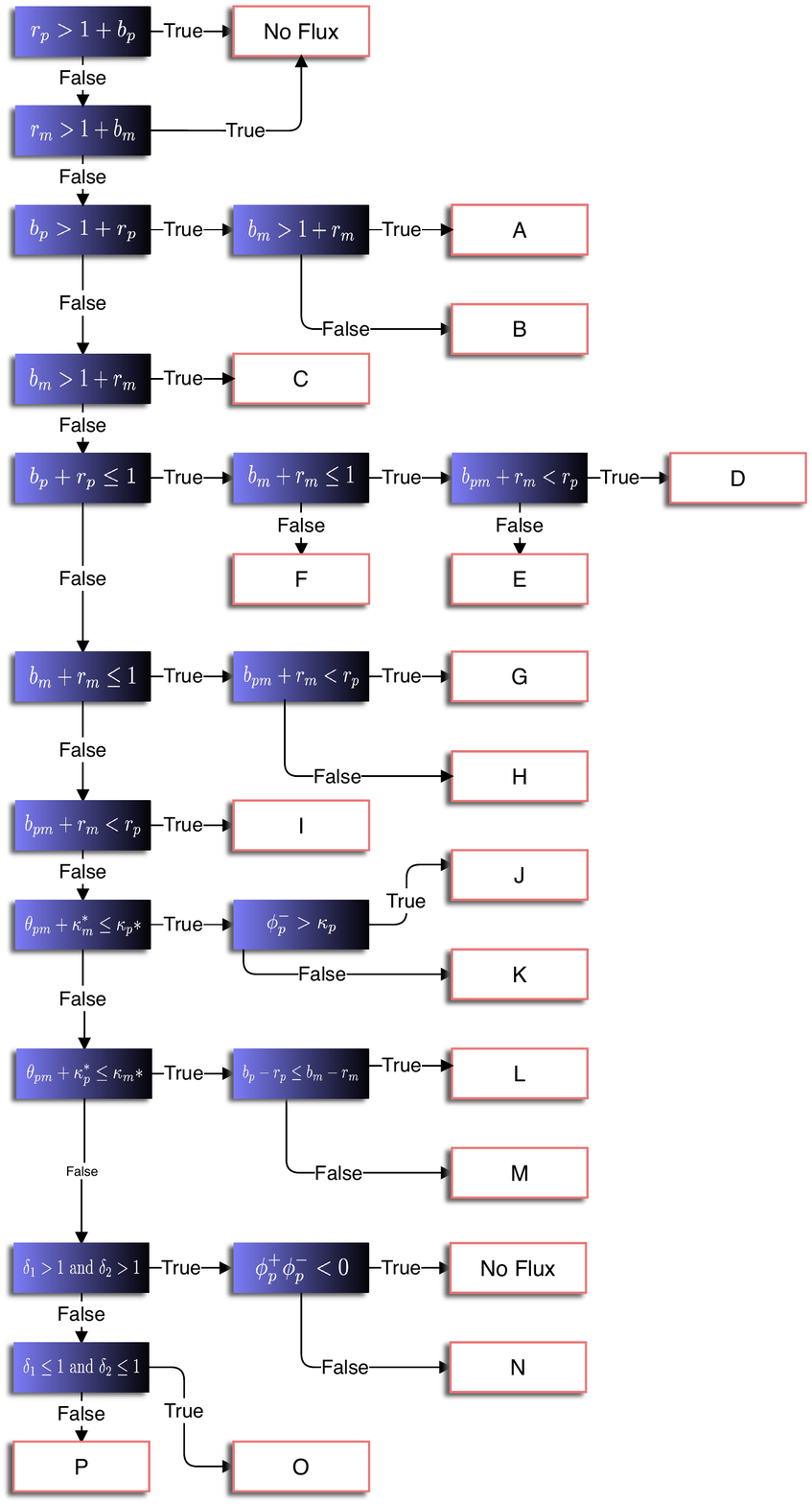}
                \caption{Decision tree when the planet and moon overlap each other ($b_{pm} < r_p + r_m$). Blue cells represent a test on the state of the system, the outcomes of which determine the cases represented by white boxes. The letters in the white boxes refer to the configurations shown in Figure \ref{fig:configurations}, and the arcs to integrate and their integration limits for each case are listed in Table \ref{tbl:configurations}. The boxes labeled ``No Flux'' indicate cases where the star is completely obscured by one or both of the transiting bodies. These outcomes are only possible when either $r_p$, $r_m$, or both are greater than 1, and thus would likely only occur for eclipsing binary or triple-star systems.}
                \label{fig:flowchart}
            \end{figure*}

    The organization of this paper is as follows: in \S \ref{sec:approach} we outline our approach to computing the integral of the limb-darkening profile over the area of the occulting bodies. This follows closely the approach of \cite{pal2012}, \cite{Luger2019}, and \cite{Agol2020}. We also show how we use the chain rule to write the derivatives with respect to the radii and positions of the two transiting bodies. In \S \ref{sec:uniform} we give the solution and its derivatives for the uniform term of the limb-darkening profile. In \S \ref{sec:linear} we give the solution and derivatives for the linear term, and in \S \ref{sec:quadratic} we do the same for the quadratic term. In \S \ref{sec:polynomial} we extend our model to polynomial limb-darkening laws. In \S \ref{sec:phi} we give formula for computing the locations of the intersections between each transiting body and the limb of the star, and the planet or moon and the star's limb. The formulae we give are crafted to be numerically stable and to mitigate roundoff error. In \S \ref{sec:limits} we show how to choose the integration limits for each arc depending on the geometry of the system at that moment. This section references the flowchart in Figure \ref{fig:flowchart} which shows the procedure for determining which geometric case the system is in at a given time. Figure \ref{fig:configurations} gives examples of the geometry for each of the cases in the flowchart, and Table \ref{tbl:configurations} gives the evaluated integral for each case in terms of the solutions for $G_N$ given in \S \ref{sec:uniform}, \S \ref{sec:linear}, \S \ref{sec:quadratic}, and \S \ref{sec:polynomial}.
    
    In \S \ref{sec:speed} we benchmark our code, both with and without the gradient computation, against \texttt{photodynam}, and in \S \ref{sec:luna} we compare the accuracy of our code with \texttt{LUNA}. Finally, in \S \ref{sec:applications} we discuss several applications of our code, focusing on conducting MCMC simulations to infer parameters of the moon/planet transit and demonstrating the use of our code by re-analysing the potential mutual transit of Kepler-51 b \& d. Lastly, we include as an appendix a minor correction to \cite{Kipping2011} and \cite{Fewell2006} that has to do with an incorrect expression for the area of overlap of three circles. 

\section{Our approach}
    \label{sec:approach}
    
    We consider the flux from a star eclipsed by two bodies, here taken to be an exoplanet and its moon, though they could also represent two exoplanets, two stars, or a combination of these. For clarity and consistency throughout the paper we refer to the two transiting bodies as the planet and moon, with the planet being the larger body and the moon the smaller. We frequently use ``$p$'' and ``$m$'' as subscripts to refer to parameters that pertain to the planet and moon (or the larger and smaller bodies) respectively. 
    
\begin{figure}
        \centering
        \includegraphics[width=\hsize]{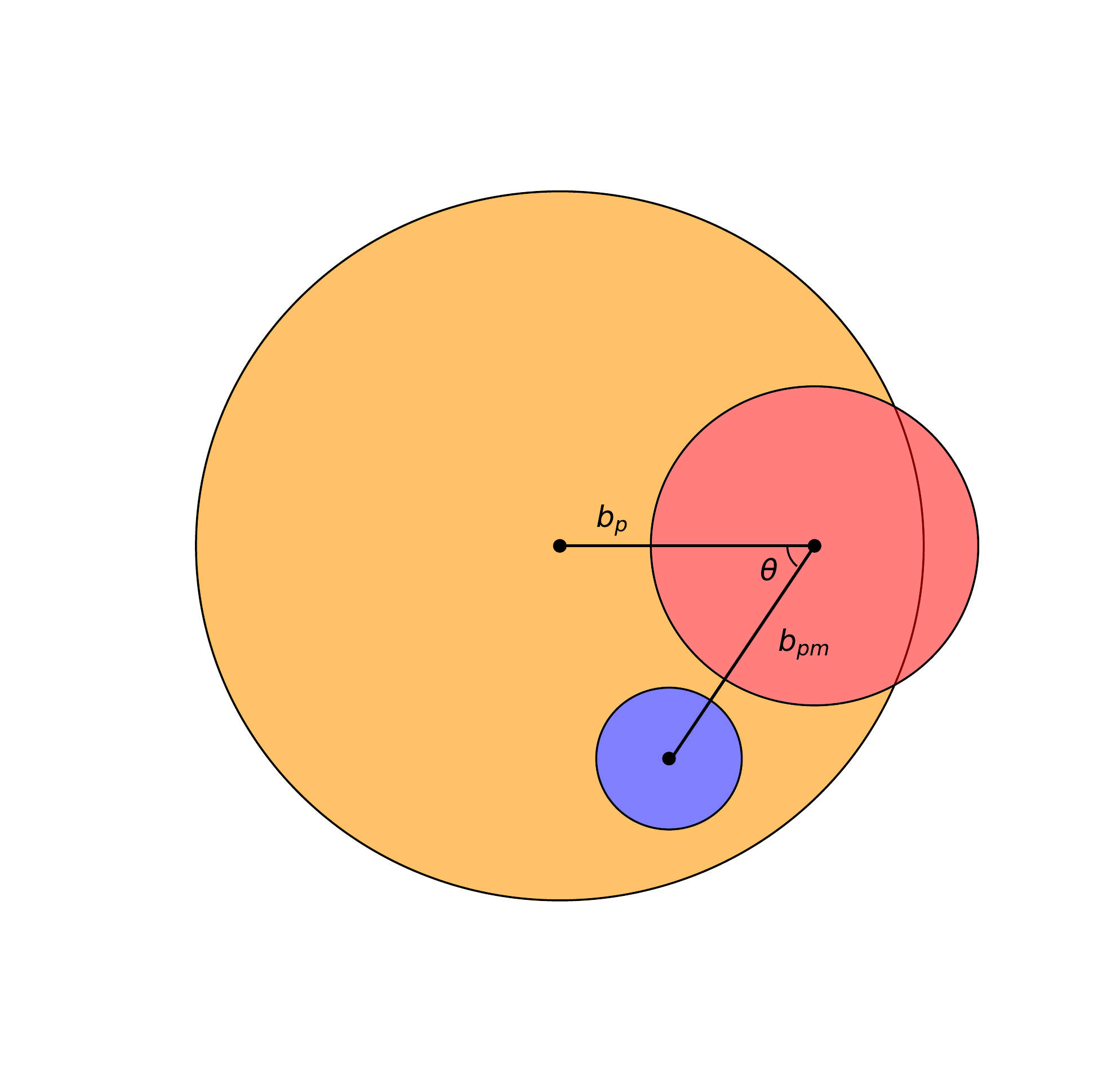}
        \caption{A sample configuration of the transiting exoplanet (red circle) and exomoon (blue circle). We use the input coordinates $\{b_p, b_{pm}, \theta\}$. Each of these coordinates are independent of the other two and together with the radii $r_p$ and $r_m$ they completely define the state of the system.}
        \label{fig:coords}
    \end{figure}

    We initially consider the star's radial surface brightness profile, $I(r)$, to be described by a quadratic limb-darkening law before expanding to consider polynomial limb-darkening. The quadratic limb-darkening profile is given by
    \begin{equation}
        \label{eqn:intensity}
        I(r) = 1 - c_1(1 - \mu) - c_2(1 - \mu)^2,
    \end{equation}
    where $\mu = \sqrt{1 - r^2}$ with $r$ being the radial distance from the center of the star projected onto the sky plane. This intensity profile can be rearranged as follows: 
    \begin{eqnarray}
        \label{eqn:quad_ld_profile}
        \nonumber I(r) &=& (1 - c_1 - 2c_2) + (c_1 + 2c_2)\mu + c_2(x^2 + y^2)\\
        &=& (1 - c_1 - 2c_2)g_0 + (c_1 + 2c_2)g_1 + c_2g_2\\
        \nonumber &=& u_0g_0 + u_1g_1 + u_2g_2,
    \end{eqnarray} 
    where we have defined $g_0 = 1$, $g_1 = \mu = \sqrt{1 - x^2 - y^2}$ and $g_2 = r^2 = x^2 + y^2$. We consider each term of this equation separately, referring to the $g_0$ term as the constant or uniform term, $g_1$ as the linear term, and $g_2$ as the quadratic term. We define a coordinate system in which the planet is centered on the positive $x$-axis at $x = b_p$ and the moon is a distance $b_{pm}$ from the planet. The line connecting the centers of the moon and planet is at an angle $\theta$ from the $x$-axis as shown in Figure $\ref{fig:coords}$.
    
    Following the approach of \cite{pal2012}, we apply Green's theorem to transform the integral from a surface integral over the eclipsed region of the star into a line integral over a vector field along the closed loop bounding the unocculted region. The integration along the boundary is split into arcs along the boundaries of the star, moon, and planet. An example configuration is shown in figure \ref{fig:coords}. 
    
    We consider the arcs belonging to the planet and the moon separately. The path along the arc of the planet is parameterized as
    \begin{eqnarray}
        x &=& b_p - r_p \cos\phi', \\
        y &=& r_p \sin\phi',
    \end{eqnarray}
    where $\phi'$ is the angle measured from the negative $x$-axis. 
    
    Once we have completed the integration along the arcs belonging to the planet, we rotate our coordinate system through an angle $\theta_{pm}$, where $\theta_{pm}$ is the angle between the center of the planet and the center of the moon relative to the center of the star, so that the moon sits on the positive x-axis at $(0, b_m)$. The path along an arc belonging to the moon can now be parameterized as 
    \begin{eqnarray}
        x &=& b_m - r_m \cos\phi', \\
        y &=& r_m \sin\phi'.
    \end{eqnarray}
    Green's theorem does not technically allow us to transform coordinate systems mid-way through the integration, as that breaks the closed loop to which the theorem applies. Fortunately, for radial limb-darkening laws it is possible to choose a vector field for Green's theorem which is invariant to rotations about the center of the star so that the integral along the arc of a circle centered at $(b, 0)$ is identical to the integral along the arc of a circle centered at $(b\cos\theta_{pm}, b\sin\theta_{pm})$ for any value of $\theta_{pm}$. 
    
    For each term in Equation \eqref{eqn:quad_ld_profile} we define the \textit{primitive integral}
    \begin{equation}
        G_n(\phi, r, b) = -\int_0^\phi \left[f_{n, y}(x, y)c_{\phi'} + f_{n, x}(x, y)s_{\phi'}\right]r d\phi',
    \end{equation}
    where $c_{\phi'}$ and $s_{\phi'}$ stand in for $\cos\phi'$ and $\sin\phi'$ respectively, and where $\bvec{f}_n$ is a vector field chosen such that
    \begin{equation}
        \frac{df_{n, y}}{dx} - \frac{df_{n, x}}{dy} = g_n.
    \end{equation} Table \ref{tbl:configurations} shows how these integrals can be combined to compute the total flux. As discussed above, we also require that the integrand be invariant with respect to rotation about the origin as this condition allows us to transform between coordinate systems where each body is positioned on the positive x-axis. While this isn't strictly necessary (we could complete the integration in a single coordinate system), it simplifies the algebra significantly and reduces the required number of computations by allowing us to compute an integral of the same form for every arc. The vector fields $\bvec{f}_0$ and $\bvec{f}_1$ are given by \cite{pal2012} as
    \begin{eqnarray}
        \bvec{f}_0 &=& \frac{1}{2}(-y\bvec{\hat{x}}+x\bvec{\hat{y}}), \\ \nonumber
        \bvec{f}_1 &=& \frac{1-(1-x^2-y^2)^{3/2}}{3(x^2+y^2)}(-y\bvec{\hat{x}}+x\bvec{\hat{y}}).
    \end{eqnarray}
    These fields meet our requirement of being invariant to rotations about the center of the star. However the expression given by \cite{pal2012} for the quadratic term does not meet this requirement. Therefore we adopt the following for $\bvec{f}_2$:
    \begin{equation}
        \bvec{f}_2 = \frac{1}{4}(x^2+y^2)(-y\bvec{\hat{x}}+x\bvec{\hat{y}}).
    \end{equation}
    Finally, for the polynomial terms we adopt the fields given by \cite{Agol2020}, which already meet our requirement:
    \begin{equation}
        \bvec{f}_n = (1-x^2-y^2)^{n/2}(-y\bvec{\hat{x}}+x\bvec{\hat{y}}),
    \end{equation} for $n\geq2$.
    
    The angle $\phi$ is the endpoint of the arc as measured from the negative $x$-axis. This angle depends on the geometry of the system and can correspond to either an intersection between the planet and moon, or an intersection between the planet or moon and the limb of the star. Thus $\phi$ is a function of $r_p$, $r_m$, $b_p$, and $b_m$. With this dependence made explicit, $G_n$ becomes a function of $r_p$, $r_m$, $b_p$, and $b_m$ as well. We can now use the chain rule to write the derivatives of $G_n$ with respect to each of these variables:
    
    \begin{eqnarray}
        \label{eqn:dgdrp}
        \nonumber \frac{dG_n(\phi, b_p, r_p)}{dr_p} &=& \frac{\partial G_n}{\partial r_p} + \frac{\partial G_n}{\partial \phi}\frac{\partial \phi}{\partial r_p}, \\
        \frac{dG_n(\phi, b_m, r_m)}{dr_p} &=& \frac{\partial G_n}{\partial \phi}\frac{\partial \phi}{\partial r_p},
    \end{eqnarray}
    \begin{eqnarray}
        \label{eqn:dgdrm}
        \frac{dG_n(\phi, b_p, r_p)}{dr_m} &=& \frac{\partial G_n}{\partial \phi}\frac{\partial \phi}{\partial r_m}, \\
        \nonumber \frac{dG_n(\phi, b_m, r_m)}{dr_m} &=& \frac{\partial G_n}{\partial r_m} + \frac{\partial G_n}{\partial \phi}\frac{\partial \phi}{\partial r_m},
    \end{eqnarray}
    \begin{eqnarray}
        \label{eqn:dgdbp}
        \nonumber \frac{dG_n(\phi, b_p, r_p)}{db_p} &=& \frac{\partial G_n}{\partial b_p} + \frac{\partial G_n}{\partial \phi}\frac{\partial \phi}{\partial b_p}, \\
        \frac{\partial G_n(\phi, b_m, r_m)}{\partial b_p} &=& \frac{\partial G_n}{\partial \phi}\frac{\partial \phi}{\partial b_p},
    \end{eqnarray}
    \begin{eqnarray}
        \label{eqn:dgdbm}
        \frac{\partial G_n(\phi, b_p, r_p)}{\partial b_m} &=& \frac{\partial G_n}{\partial \phi}\frac{\partial \phi}{\partial b_m}, \\
        \nonumber \frac{\partial G_n(\phi, b_m, r_m)}{\partial b_m} &=& \frac{\partial G_n}{\partial b_m} + \frac{\partial G_n}{\partial \phi}\frac{\partial \phi}{\partial b_m}.
    \end{eqnarray}
    
    Because we locate the moon using the parameters $b_{pm}$ and $\theta$ rather than using $b_m$ directly, we now need to show how the derivatives with respect to the set of input coordinates $\{b_p, b_{pm}, \theta\}$ are recovered from the integrals with respect to $b_p$ and $b_m$ using the chain rule. For the derivatives of $G_n$ we have:
    \begin{eqnarray}
        \nonumber \frac{dG_n(\phi, b_m, r_m)}{db_p} &=& \frac{\partial G_n}{\partial b_p} + \frac{\partial G_n}{\partial b_m}\frac{\partial b_m}{\partial b_p}, \\ 
        \nonumber \frac{dG_n(\phi, b_m, r_m)}{db_{pm}} &=& \frac{\partial G_n}{\partial b_m}\frac{\partial b_m}{\partial b_p}, \\
        \frac{dG_n(\phi, b_p, r_p)}{db_{pm}} &=& \frac{dG_n}{d\phi}\frac{\partial \phi}{\partial b_{pm}},\\
        \nonumber \frac{dG_n(\phi, b_m, r_m)}{d\theta} &=& \frac{\partial G_n}{\partial b_m}\frac{\partial b_m}{\partial \theta}, \\
        \nonumber \frac{dG_n(\phi, b_p, r_p)}{d\theta} &=& \frac{\partial G_n}{\partial b_m}\frac{\partial b_m}{\partial \theta}.
    \end{eqnarray}
    
    The derivatives of the angle $\phi$ which appear in Equations \eqref{eqn:dgdrp}, \eqref{eqn:dgdrm}, \eqref{eqn:dgdbp}, and \eqref{eqn:dgdbm} will be dealt with in section \ref{sec:phi}. The derivatives of $b_m$ are
    \begin{eqnarray}
        \nonumber \frac{\partial b_m}{\partial \theta} &=& \frac{b_pb_{pm}\sin\theta}{b_m},\\
        \frac{\partial b_m}{\partial b_{pm}} &=& \frac{b_{pm} - b_p\cos\theta}{b_m},\\
        \nonumber \frac{\partial b_m}{\partial b_p} &=& \frac{b_p - b_{pm}\cos\theta}{b_m},
    \end{eqnarray}
    where $b_m$ is computed using the formula
    \begin{equation}
        b_m = \sqrt{(b_p-b_{pm})^2 + 2b_pb_{pm}(1 - \cos\theta)},
    \end{equation}
    which is a re-arranged version of the familiar law of cosines constructed to eliminate round-off error \citep{kahan2000}. 

\section{Evaluating the Primitive Integrals}

With the description of the primitive integral complete, we now evaluate the integral and its derivatives for each value of $n$, starting with $n=0$; i.e.\ a uniformly emitting source.

    \subsection{Uniform Limb-Darkening}
    \label{sec:uniform}

    For the uniform term the primitive integral is 
    \begin{equation}
        \label{eqn:prim_uniform}
        G_0(\phi, r, b) = \int_0^\phi \frac{r}{2}(r - b\cos\phi')d\phi',
    \end{equation}
    which gives us 
    \begin{equation}
        G_0(\phi, r, b) = \frac{r}{2}(r\phi - b\sin\phi).
    \end{equation}
    This has the derivatives
    \begin{eqnarray}
        \frac{\partial G_0}{\partial r} &=& r\phi - \frac{b}{2}\sin\phi,\\
        \frac{\partial G_0}{\partial b} &=& -\frac{r}{2}\sin\phi,\\
        \frac{\partial G_0}{\partial \phi} &=& \frac{r}{2}(r - b\cos\phi).
    \end{eqnarray}

\subsection{Linear Limb-Darkening}
    \label{sec:linear}
    
    For the linear term ($n=1$) the primitive integral is 
    \begin{equation}
        \label{eqn:prim_linear}
        G_1(\phi, r, b) = \frac{r}{3}\int_{0}^{\phi}\frac{1 - (\mu^2 - b^2+2brc_{\phi'})^{3/2}}{b^2+r^2-2brc_{\phi'}}(r - bc_{\phi'})d\phi'.
    \end{equation}
    When this integral is evaluated it yields a combination of incomplete elliptic integrals of the first, second, and third kind which we express in the form:
    \begin{eqnarray}
        \label{eqn:general_integral}
        \nonumber G_1(\phi, r, b) &=& \alpha(r, b)E(s, k) + \beta(r, b)F(s, k) \\ &+& \gamma(r, b)\Pi(s, n, k) + R(\phi, r, b),
    \end{eqnarray}
    where $s = s(\phi)$ is the amplitude of the incomplete elliptic integrals, $n = n(r, b)$ is the parameter, and $k = k(r, b)$ is the modulus, following the conventions of \cite{Byrd1954}. The functions $\alpha(r, b)$, $\beta(r, b)$ and $R(\phi, r, b)$ are defined as
    
    
        \begin{equation}
            \label{eqn:alpha}
            \alpha(r, b) = 
            \begin{cases}
                \frac{1}{9}\sqrt{1-(b-r)^2}\ (7r^2+b^2-4) & b + r \leq 1 \\
                \frac{2}{9}\sqrt{br}\ (7r^2+b^2-4) & b + r > 1 ,
            \end{cases}
        \end{equation}
        \begin{equation}
            \label{eqn:beta}
            \beta(r, b) = 
            \begin{cases} 
                \frac{r^4+b^4+r^2-b^2(5+2r^2)+1}{9\sqrt{1-(b-r)^2}} & b + r \leq 1 \\
                -\frac{3 + 2r(b^3 + 5b^2r+3r(r^2-2) + b(7r^2-4))}{18\sqrt{br}} & b + r > 1,
            \end{cases}
        \end{equation}
        \begin{equation}
            \label{eqn:gamma}
            \gamma(r, b) = 
            \begin{cases}  
                \frac{b+r}{3(b-r)}\frac{1}{\sqrt{1-(b-r)^2}} & b + r \leq 1 \\
                \frac{b+r}{6(b-r)\sqrt{br}} & b + r > 1,
            \end{cases}
        \end{equation} and
        \begin{eqnarray}
            \label{eqn:R}
            R(\phi, r , b) = 
                \frac{\phi}{6} &-&          \frac{1}{3}\tan^{-1}\left(\frac{b+r}{b-r}\tan\left(\frac{\phi}{2}\right)\right) \\ \nonumber &-& \frac{2br}{9}\sin\phi\sqrt{1-b^2-r^2+2br\cos\phi}.
        \end{eqnarray}
        The formula for $R(\phi, r, b)$ is the same in the $b+r>1$ case as in the $b+r<1$ case.

        The amplitude, characteristic, and parameter of the incomplete elliptic integrals is given by:
        
        \begin{equation}
            \label{eqn:s}
            s = 
            \begin{cases}
                \frac{\phi}{2} & b + r \leq 1 \\
                \sin^{-1}\left(\frac{2\sqrt{br}\sin(\phi/2)}{\sqrt{1 - (r-b)^2}}\right) & b + r > 1
            \end{cases},
        \end{equation}
        
        \begin{equation}
            \label{eqn:m}
            k^2 = 
            \begin{cases}
                   \frac{4br}{1-(r-b)^2} & b + r \leq 1 \\
                   \frac{1 - (r - b)^2}{4br} & b + r > 1
            \end{cases},
        \end{equation}
        and
        \begin{equation}
            \label{eqn:n}
            n = 
            \begin{cases}
                -\frac{4rb}{(b-r)^2} & b + r \leq 1 \\
                1 - \frac{1}{(b-r)^2} & b + r > 1
            \end{cases}.
        \end{equation}
        
        When $b=0$ and $r=1$ the solution simplifies to $G_1(\phi, b, r) = \frac{\phi}{3}$, which is useful when integrating along the boundary of the star.
        
        When $\phi$ corresponds to the intersection between the arc and the limb of the star, the amplitude of the incomplete elliptic integrals becomes $\pi/2$ and the solution can be computed with complete rather than incomplete elliptic integrals. In this case the integral evaluates to
        \begin{eqnarray}
            \nonumber G_1(\phi, r, b) = &\alpha&(r, b)E(k) + \beta(r, b)F(k) + \gamma(r, b)\Pi(n, k) \\ &+& \frac{\phi}{6} -          \frac{1}{3}\tan^{-1}\left(\frac{b+r}{b-r}\tan\left(\frac{\phi}{2}\right)\right),
        \end{eqnarray}
        where $\alpha$, $\beta$, and $\gamma$ are given by the expressions for the $b + r > 1$ case and $\phi$ is the angle to the intersection of the arc with the star's limb.
        
        \subsubsection{Derivatives for the Linear Limb-Darkening Case}
        The partial derivatives of $G_1(\phi, r, b)$ with respect to $b$ and $r$ are 
        \begin{eqnarray}
            \label{eqn:partials}
            \frac{\partial G_1}{\partial r} &=& u_r(r, b)E(s, k) + v_r(r, b)F(s, k) + p_r(\phi, r, b),\\
            \nonumber \frac{\partial G_1}{\partial b} &=& u_b(r, b)E(s, k) + v_b(r, b)F(s, k) + p_b(\phi, r, b).
        \end{eqnarray}
        
        The coefficients by which the elliptic integrals are multiplied are given by
        \begin{equation}
            u_r(r, b) = \begin{cases}
                2r\sqrt{1 - (b-r)^2} & r + b \leq 1 \\
                4\sqrt{br^3} & r + b > 1
            \end{cases},
        \end{equation}
        \begin{equation}
            v_r(r, b) = \begin{cases}
                0 & b + r \leq 1 \\
                -\sqrt\frac{r}{b}((b + r)^2 - 1) & b + r < 1
            \end{cases},
        \end{equation}
        \begin{equation}
            u_b(r, b) = \begin{cases}
                \frac{1 - (b-r)^2}{3b}(b^2 + r^2 - 1) & b + r \leq 1 \\
                \frac{2}{3}\sqrt{\frac{r}{b}}(b^2 + r^2 - 1) & b + r > 1
            \end{cases},
        \end{equation}
        \begin{equation}
            v_b(r, b) = \begin{cases}
                \frac{b^4 - (r^2-1)^2 - 2b^2(r^2+1)}{3b\sqrt{1 - (b-r)^2}} & b + r \leq 1 \\
                -\frac{1}{3}\sqrt\frac{r}{b}((b + r)^2 - 1) & b + r > 1
            \end{cases},
        \end{equation}
        and the functions $p_r(\phi, r, b)$ and $p_b(\phi, r, b)$ are given by
        \begin{eqnarray*}
            p_r(\phi, r, b) &=& \frac{b\sin\phi}{3\xi}\left((1-\xi)^{3/2}-1\right)
            \\
            p_b(\phi, r, b) &=& \frac{r\sin\phi}{3\xi}\left(1-(1+\xi)\sqrt{1-\xi}\right)
        \end{eqnarray*}
        where $\xi= b^2+r^2-2br\cos{\phi}$.
        
        The partial derivative with respect to $\phi$ is 
        \begin{equation}
            \frac{\partial G_1(\phi, r, b)}{\partial \phi} = \frac{1 - (1 - \xi)^{3/2}}{\xi}(r^2 - br\cos\phi).
        \end{equation}
        
    \subsection{Quadratic Limb-darkening}
        \label{sec:quadratic}
        For the quadratic term ($n=2$), the primitive integral is 
        \begin{eqnarray}
            \label{eqn:prim_quad}
            G_2(\phi, r, b) = \int_0^\phi &\frac{r}{4}&\left[\right.r^3 + b^2r(2 + \cos(2\phi')) \\ \nonumber &-& b(b^2+3r^2)c_{\phi'}\left.\right]d\phi',
        \end{eqnarray}
        which evaluates to 
        \begin{eqnarray}
            G_2(\phi, r, b) = &\frac{r}{4}&\left[\right.(2b^2r + r^3)\phi \\ \nonumber&+& (b^2r\cos\phi - 3br^2-b^3)\sin\phi\left.\right].
        \end{eqnarray}
        This has the derivatives
        \begin{eqnarray}
            \nonumber \frac{\partial G_2}{\partial r} &=& \frac{b}{4}(b^2 + 9r^2-2br\cos\phi)\sin\phi + rb^2+r^2,\\
            \frac{\partial G_2}{\partial b} &=& \frac{r}{4}br(4\phi + \sin(2\phi) - 3(b^2+r^2)\sin\phi),\\
            \nonumber \frac{\partial G_2}{\partial \phi} &=& \frac{r}{4}\left(2b^2r + r^3 + b(br\cos(2\phi) - (b^2+3r^2)\cos\phi)\right).
        \end{eqnarray}
        
    \subsection{Polynomial Limb-darkening}
        \label{sec:polynomial}
        We now give the general solution for polynomial limb-darkening laws of the form
        \begin{equation}
            \label{eqn:poly_ld_profile}
            I(r) = 1 - \sum_{i=1}^N c_i(1-\mu)^i,
        \end{equation}
        where again $\mu = \sqrt{1 - r^2} = \sqrt{1 - x^2-y^2}$. We follow \cite{Agol2020} in writing this limb-darkening law in Green's basis as 
        \begin{equation}
            I(r) = \mathbf{g}^\mathrm{T} \mathbf{u},
        \end{equation}
        where $\mathbf{g}$ is defined
        \begin{equation}
            \label{eqn:greens}
            \mathbf{g} = (1, \mu, 4\mu^2-2, 5\mu^3-3\mu, \dots, (N+2)\mu^N - N\mu^{N-2})^\mathrm{T},
        \end{equation}
        and $\mathbf{u}$ is a vector with elements defined recursively by
        \begin{equation}
            u_n = \begin{cases}
                \frac{p_n}{n+2} + u_{n+2} & N\geq n\geq 2 \\
                p_n + (n + 2)u_{n+2} & n = 1, 0
            \end{cases},
        \end{equation}
        with $u_{N+1}=g_{N+2}=0$, and
        where the elements of $\mathbf{p}$ are related to the elements of $\mathbf{c} = (c_0, c_1, c_2, \dots, c_N)$ by:
        \begin{equation}
            p_n = (-1)^{n+1}\sum_{m=0}^n {m \choose n}c_m,
        \end{equation}
        with $c_0 = -1$.
        It is important to note that while this basis shares the first two terms with Equation \eqref{eqn:intensity}, the quadratic and higher order terms are distinct. Therefore the solution for the quadratic term given above should not be re-used when computing light curves for higher order limb-darkening laws. 
        
        Our goal is now to compute the integral over each term in Equation \eqref{eqn:greens} along an arbitrary arc. The appropriate primitive integral in this case is 
        \begin{equation}
            \label{eqn:prim_poly}
            G_n(\phi, r, b) = \int_0^\phi r(r - bc_{\phi'})(1 - r^2 - b^2 + 2brc_{\phi'})^{n/2}d\phi',
        \end{equation}
        for $n \geq 2$, which we can rewrite in the form 
        \begin{equation}
            \label{eqn:polynomial_ld}
            G_n(\phi, r, b) = (1 + r^2 - b^2)\mathcal{M}_n - \mathcal{M}_{n+2},
        \end{equation}
        where
        \begin{equation}
            \mathcal{M}_n = \int_0^\phi \frac{1}{2}(1 - r^2 - b^2 + 2br\cos\phi^\prime)^{n/2}d\phi^\prime.
        \end{equation}
        This integral obeys the recursion relation 
        \begin{eqnarray}
            n\mathcal{M}_n &=& 2(n-1)(1-b^2-r^2)\mathcal{M}_{n-2} \\ \nonumber &+& (n-2)(1 - (b-r)^2)((b+r)^2-1)\mathcal{M}_{n-4} \\ \nonumber &+& 2br\sin{\phi}(1-b^2-r^2+2br\cos\phi)^{\frac{n}{2}-1}.
        \end{eqnarray}
        Given the $\mathcal{M}_1$ through $\mathcal{M}_4$ we can then use this recursion relation to compute $\mathcal{M}_n$ for any $n$, which can then be plugged into Equation \eqref{eqn:polynomial_ld} to find the integral over any term in $\mathbf{g}$.
        
        We can re-use $\mathcal{M}_n$ to write the derivatives of the polynomial terms:
        
        \begin{eqnarray}
            \frac{\partial G_n}{\partial r} = &2&r\left[(2+n)\mathcal{M}_n-n\mathcal{M}_{n-2}\right]\\\nonumber&-&b\sin{\phi}(1-b^2-r^2+2br\cos\phi)^{\frac{n}{2}},
        \end{eqnarray}
        \begin{eqnarray}
            \nonumber \frac{\partial G_n}{\partial b} = &\frac{n}{b}&\left[(r^2+b^2)(\mathcal{M}_n-\mathcal{M}_{n-2}) + (r^2-b^2)^2\mathcal{M}_{n-2}\right]\\\nonumber&-&r\sin{\phi}                         (1-b^2-r^2+2br\cos\phi)^{\frac{n}{2}},
        \end{eqnarray}
        and
        \begin{eqnarray}
            \nonumber\frac{\partial G_n}{\partial \phi} &=& r(r-b\cos\phi)(1-r^2-b^2+2br\cos\phi)^\frac{n}{2}.
        \end{eqnarray}
        
        Finally, we give the first four terms of the recursion which are necessary to compute subsequent terms:
        \begin{eqnarray}
            \mathcal{M}_0 &=& \frac{\phi}{2},\\
            \nonumber \mathcal{M}_1 &=& \alpha_1 E(s, k) + \beta_1 F(s,k),\\
            \nonumber \mathcal{M}_2 &=& \frac{1}{2}(1-b^2-r^2)\phi + br\sin\phi,\\
            \nonumber \mathcal{M}_3 &=& \alpha_3 E(s, k) + \beta_3 F(s, k) + R_3(\phi, r, b),
        \end{eqnarray}
        
        where the coefficients for $\mathcal{M}_1$ are
        \begin{eqnarray}
            \alpha_1 &=& \begin{cases}
                \sqrt{1-(b-r)^2}& b + r \leq 1 \\
                2\sqrt{br}& b + r > 1
            \end{cases},\\
            \nonumber \beta_1 &=& \begin{cases}
                0 & b+r \leq 1 \\
                \frac{1-(b+r)^2}{2\sqrt{br}} & b+r > 1
            \end{cases},
        \end{eqnarray}
        and for $\mathcal{M}_3$,
        \begin{eqnarray}
            \alpha_3 &=& \begin{cases}
                \frac{4}{3}(1-b^2-r^2)\sqrt{1-(b-r)^2} & b+r \leq 1 \\
                \frac{8}{3}(1-b^2-r^2)\sqrt{br} & b + r > 1
            \end{cases},\\
            \nonumber \beta_3 &=& \begin{cases}
                 \frac{2r^2(b^2+1)-(b^2-1)^2-r^4}{3\sqrt{1-(b-r)^2}} & b+r \leq 1 \\
                \frac{1-(b+r)^2}{6\sqrt{br}}(3(1-b^2-r^2)-2br) & b+r > 1
            \end{cases}.
        \end{eqnarray}
        The function $R_3$ is given by
        \begin{equation}
            R_3(\phi, r, b) = \frac{2}{3}br\sin\phi\sqrt{1-b^2-r^2+2br\cos\phi}.
        \end{equation}
        
        The definitions of $k$ and $s$ are the same as for the linear limb-darkening case and are given in Equations \eqref{eqn:s} and \eqref{eqn:m}.
        
        We do not yet include polynomial limb-darkening laws in \texttt{gefera}, as in most cases quadratic limb-darkening will be sufficiently accurate to model an observed light curve. As a result, we have not tested these equations for numerical stability to the extent that we have for uniform, linear, and quadratic expressions. Anyone interested in implementing this solution should therefore take care to check for stability and accuracy. Interested readers might find it useful to refer back to \cite{Agol2020} as a guide. 
        
    \section{Computing $\phi$}
        \label{sec:phi}
    
        The angle $\phi$ is the one-sided integration limit as an angle along the arc, measured from the negative $x$-axis. It can take on values in the range $(-\pi, \pi)$. For $\phi < 0$ the integrals can be computed using the property $G(-\phi, r, b) = G(\phi, r, b)$. Depending on the geometry of the system, $\phi$ will represent either the intersection of the planet and moon, or the intersection of the planet or moon with the limb of the star. In the case that body we're integrating along the arc of does not intersect either the other body or the limb of the star, we will have $\phi = \pm\pi$. In this section we show how to compute $\phi$ when the endpoint of the arc is an intersection with the other transiting body or with the stellar limb.
        
        \subsection{Intersections between planet and moon}
\begin{figure}
                \includegraphics[width=\hsize]{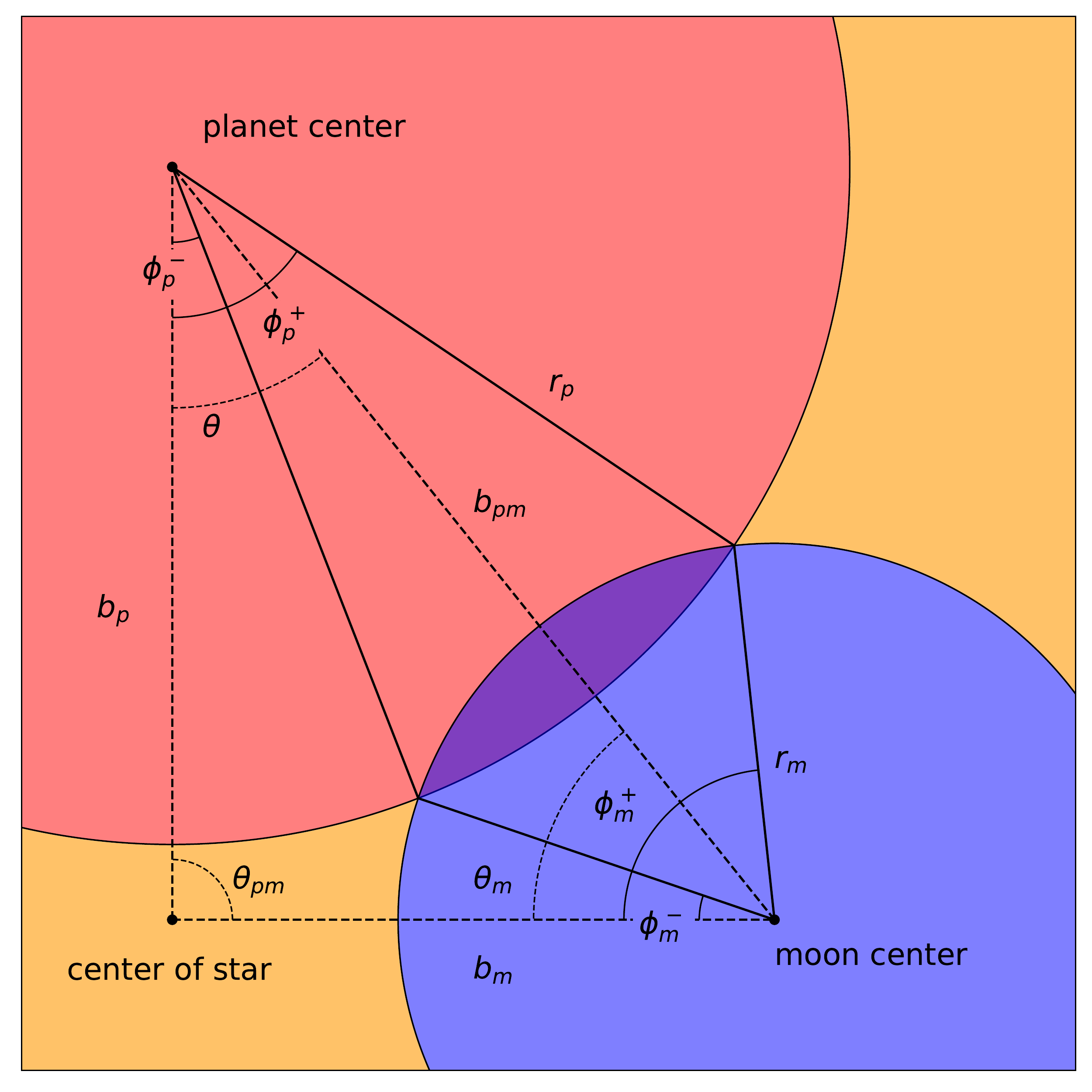}
                \caption{Geometry of the planet-moon overlap showing the angles $\phi_p^\pm$ and $\phi_m^\pm$ from Equations \eqref{eqn:phip} and $\eqref{eqn:phim}$. The dashed lines show the triangle formed by the sides $b_p$, $b_m$ and $b_{pm}$.}
                \label{fig:planet_moon_overlap}
            \end{figure}

            When the planet and moon overlap we need to compute four angles, these being the angles to the two points of intersection with respect to the centers of each body. Figure \ref{fig:planet_moon_overlap} shows the geometry of the planet/moon overlap. The two points of intersection with respect to the planet are measured from the line connecting the center of the star to the center of the planet. These are given by 
            \begin{eqnarray}
                \label{eqn:phip}
                \nonumber \phi^+_p &=& \theta + \phi_p, \\
                \phi^-_p &=& \theta - \phi_p,
            \end{eqnarray}
            Similarly, the two points of intersection with respect to the moon are measured from the line connecting the center of the star to the center of the moon and are given by 
            \begin{eqnarray}
                \label{eqn:phim}
                \nonumber \phi^+_m &=& \theta_m + \phi_m,\\
                \phi^-_m &=& \theta_m - \phi_m,
            \end{eqnarray}
            where $\theta_m$ is the angle between the line connecting the centers of the planet and moon and the line connecting the centers of the star and moon, as seen in Figure \ref{fig:planet_moon_overlap}. This angle is computed as 
            \begin{equation}
                \theta_m = \mathrm{Atan2}(b_p\sin\theta, b_{pm} - b_p\cos\theta).\footnote{The function Atan2$(y, x)$ is equivalent to Atan$(y/x)$. It is available as a built-in math function in most programming languages. We have found that this alternative formulation of the arctangent function is more numerically stable in the limit $y\gg x$.}
            \end{equation}
            
            We use the method outlined in \cite{Agol2020} to compute $\phi_p$ and $\phi_m$ to high accuracy. First we compute a quantity we call $\Delta$, which is four times the area of the triangle formed by $r_p$, $r_m$, and $b_{pm}$, using the modified version of Heron's formula by \cite{kahan2000}:
            \begin{equation}
                \label{eqn:heron}
                \Delta = \sqrt{(a + (b + c))(c - (a - b))(c + (a - b))(a + (b - c))},
            \end{equation}
            where $a$, $b$ and $c$ are $r_p$, $r_m$, and $b_{pm}$ ordered from highest to lowest. In this formula the exact placement of the parentheses is essential to preserving numerical accuracy and should not be altered. The angles are then given by 
            \begin{eqnarray}
                \nonumber \phi_p &=& \mathrm{Atan2}(\Delta, (r_p - r_m)(r_p+r_m)+b_{pm}^2),\\
                \phi_m &=& \mathrm{Atan2}(\Delta, (r_m - r_p)(r_m+r_p)+b_{pm}^2).
            \end{eqnarray}
            For the derivatives of $\phi_p$ and $\theta$ we have 
            \begin{eqnarray}
                \nonumber \frac{\partial\phi_p}{\partial r_p} &=& \frac{(b_{pm} - r_p) (b_{pm} + r_p) - r_m^2}{r_p\Delta},\\
                \nonumber \frac{\partial \phi_p}{\partial r_m} &=& \frac{2r_m}{\Delta},\\
                \nonumber \frac{\partial \phi_p}{\partial b_{pm}} &=& \frac{(r_p - r_m)(r_p + r_m) - b_{pm}^2}{b_{pm}\Delta},\\
                \frac{\partial \phi_p}{\partial b_p} &=& \frac{\partial \phi_p}{\partial \theta} = 0.
            \end{eqnarray}
            The derivatives of $\theta$ are all equal to zero except for the derivative of $\theta$ with respect to itself, which is $1$. For $\phi_m$ and $\theta_m$ we have
            \begin{eqnarray}
                \nonumber \frac{\partial \phi_m}{\partial r_p} &=& \frac{2r_p}{\Delta},\\
                \nonumber \frac{\partial \phi_m}{\partial r_m} &=& \frac{(b_{pm} - r_m)(b_{pm} + r_m) - r_p^2}{r_m\Delta},\\
                \nonumber \frac{\partial \phi_m}{\partial b_{pm}} &=& \frac{(r_m + b_{pm})(r_m - b_{pm}) - r_p^2}{b_{pm}\Delta},\\
                \frac{\partial \phi_m}{\partial b_p} &=& \frac{\partial \phi_m}{\partial \theta} = 0,
                \end{eqnarray}
                \begin{eqnarray}
                \nonumber\frac{\partial \theta_m}{\partial r_p} &=& \frac{\partial \theta}{\partial r_m} = 0,\\
                \nonumber \frac{\partial \theta_m}{\partial b_p} &=& \frac{b_{pm}\sin\theta}{b_m^2},\\
                \nonumber \frac{\partial \theta_m}{\partial b_{pm}} &=&  -\frac{b_p\sin\theta}{b_m^2},\\
                \nonumber \frac{\partial \theta_m}{\partial \theta} &=& \frac{(b_{pm} - b_m)(b_{pm} + b_m) - b_p^2}{2b_m^2}.
            \end{eqnarray}
            
        \subsection{Intersections between the planet/moon and stellar limb}
        
\begin{figure}
                \centering
                \includegraphics[width=\hsize]{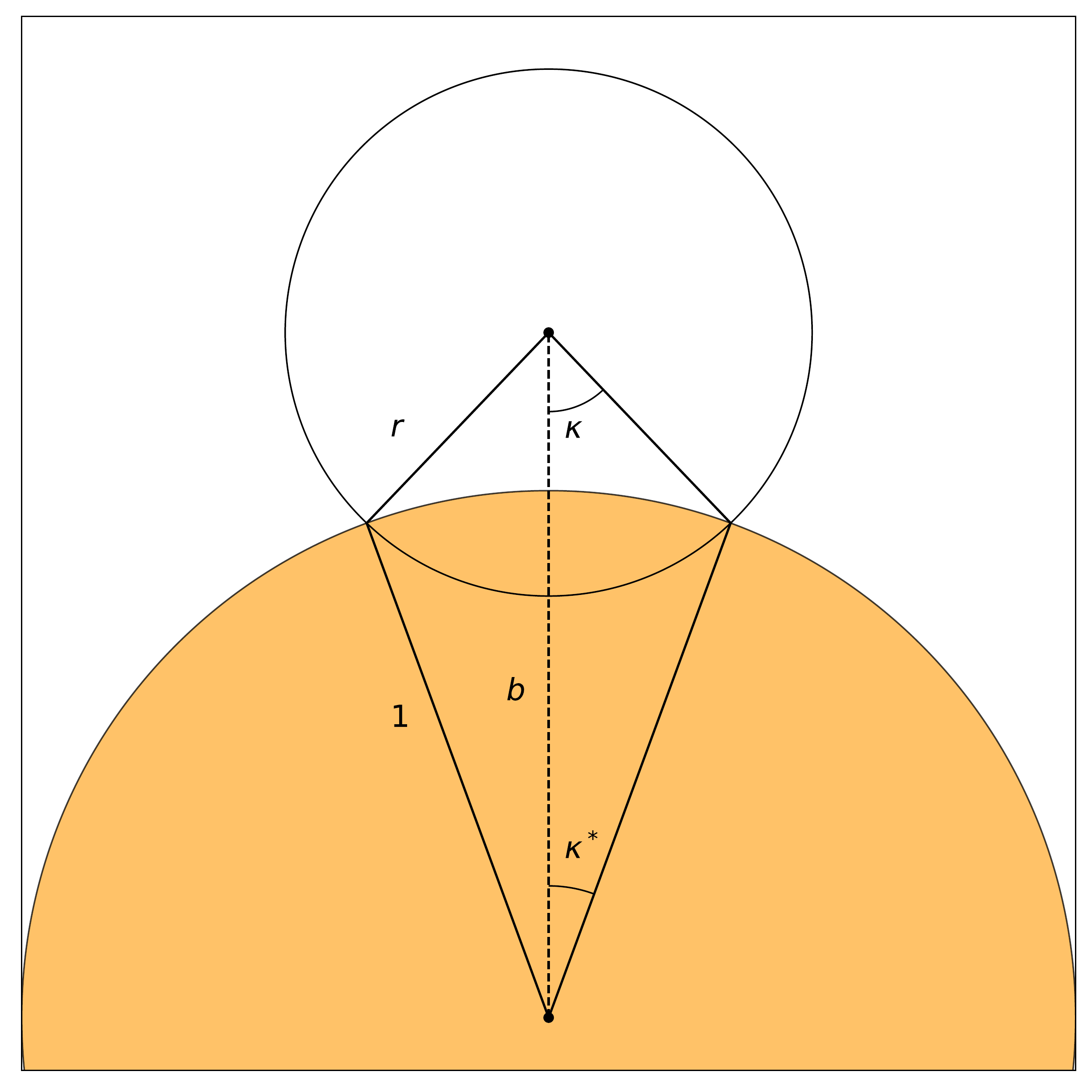}
                \caption{Geometry of the overlap between the star and a transiting planet or moon showing the angles $\kappa$ and $\kappa^*$ defined in Equations \eqref{eqn:kappa_p} and \eqref{eqn:kappa_m}. Here the filled yellow circle represents the star and the smaller un-filled circle represents either the planet or the moon. In the text $\kappa$ and $\kappa^*$ are subscripted with $m$ or $p$ for the moon and planet respectively.}
                \label{fig:star_overlap}
            \end{figure}

            When the planet or moon intersects the limb of the star we again must compute four angles. The first two are the angles to the two points of intersection with respect to the planet or moon, which are used to integrate along the arc of that body, and the second two are the points of intersection with respect to the center of the star which are used to integrate along the limb of the star. The angles to the intersections with respect to the center of the planet or moon are represented by $\kappa_p$ or $\kappa_m$, and those with respect to the center of the star are represented by $\kappa_p^*$ or $\kappa_m^*$. Figure \ref{fig:star_overlap} illustrates this geometry. All of these angles are computed relative to the line connecting the center of the star to the body in question. For the planet we have 
            \begin{eqnarray}
                \label{eqn:kappa_p}
                \nonumber \kappa_p &=& \mathrm{Atan2}(\Delta, (r_p - 1)(r_p + 1) + b_p^2) \\
                \kappa_p^* &=& \mathrm{Atan2}(\Delta, (1 - r_p)(1 + r_p) + b_p^2)
            \end{eqnarray}
            where $\Delta$ is given by equation $\eqref{eqn:heron}$ with $a$, $b$ and $c$ equal to $1$, $r_p$, and $b_p$ ordered from highest to lowest, and for the moon we have
            \begin{eqnarray}
                \label{eqn:kappa_m}
                \nonumber \kappa_m &=& \mathrm{Atan2}(\Delta, (r_m - 1)(r_m + 1) + b_m^2) \\
                \kappa_m^* &=& \mathrm{Atan2}(\Delta, (1 - r_m)(1 + r_m) + b_m^2)
            \end{eqnarray}
            where $\Delta$ is again given by \eqref{eqn:heron} but with $a$, $b$ and $c$ equal to $1$, $r_m$, and $b_m$ ordered from highest to lowest. The geometry of the intersection between star and transiting body is shown in figure \ref{fig:star_overlap}
            
            For the derivatives we have
            \begin{eqnarray}
                \nonumber \frac{\partial \kappa_p}{\partial r_p} &=& \frac{(b_p + r_p)(b_p - r_p) - 1}{r_p\Delta},\\
                \nonumber \frac{\partial \kappa_p}{\partial r_m} &=&   \frac{\partial \kappa_p}{\partial b_{pm}} = \frac{\partial\kappa_p}{\partial\theta} = 0,\\
                \frac{\partial\kappa_p}{\partial b_p} &=& \frac{(r_p + b_p)(r_p - b_p) - 1}{b_p\Delta},
            \end{eqnarray}
            \begin{eqnarray}
                \nonumber \frac{\partial\kappa_p^*}{\partial r_p} &=& \frac{2r_p}{\Delta},\\
                \nonumber \frac{\partial\kappa_p^*}{\partial r_m} &=& \frac{\partial\kappa_p^*}{\partial b_{pm}}= \frac{\partial\kappa_p^*}{\partial\theta} = 0,\\
                \frac{\partial\kappa_p^*}{\partial b_p} &=& \frac{(1 + b_p)(1 - b_p) - r_p^2}{b_p\Delta},
            \end{eqnarray}
            \begin{eqnarray}
                \nonumber \frac{\partial\kappa_m}{\partial r_p} &=& 0,\\
                \nonumber \frac{\partial\kappa_m}{\partial r_m} &=& \frac{(b_m + r_m)(b_m - r_m) - 1}{r_m\Delta},\\
                \frac{\partial\kappa_m}{\partial b_p} &=& \frac{(r_m + b_m)(r_m - b_m) - 1}{b_m^2\Delta}(b_p - b_{pm}\cos\phi),\\
                \nonumber \frac{\partial\kappa_m}{\partial b_{pm}} &=&  \frac{(r_m + b_m)(r_m - b_m) - 1}{b_m^2\Delta}(b_{pm}-b_p\cos\theta),\\
                \nonumber \frac{\partial\kappa_m}{\partial\theta} &=&  \frac{(r_m + b_m)(r_m - b_m) - 1}{b_m^2\Delta}b_pb_{pm}\sin\theta,
            \end{eqnarray}
            and
            \begin{eqnarray}
                \nonumber \frac{\partial\kappa_m^*}{\partial r_p} &=& 0,\\
                \nonumber \frac{\partial\kappa_m^*}{\partial r_m} &=& \frac{2r_m}{\Delta},\\
                \frac{\partial\kappa_m^*}{\partial b_p} &=& \frac{(1 + b_m)(1 - b_m) - r_m^2}{b_m^2\Delta}(b_p - b_{pm}\cos\phi),\\
                \nonumber \frac{\partial\kappa_m^*}{\partial b_{pm}} &=& \frac{(1 + b_m)(1 - b_m) - r_m^2}{b_m^2\Delta}(b_{pm}-b_p\cos\theta),\\
                \nonumber \frac{\partial\kappa_m^*}{\partial \theta} &=& \frac{(1 + b_m)(1 - b_m) - r_m^2}{b_m^2\Delta}b_pb_{pm}\sin\theta.
            \end{eqnarray}
            
    \section{Finding the Integration Limits}
        \label{sec:limits}
        
        \subsection{No planet-moon overlap}
            When $b_{pm} > b_p + b_m$ the planet and moon do not overlap each other. In this case we integrate counter-clockwise around the region of overlap between the transiting body and the star, and then to subtract this from the total flux of the star. For each body we determine whether or not it is entirely outside the perimeter of the star ($b > 1 + r$), overlapping the limb of the star ($1 - r < b < 1 + r$), or completely overlapping the star ($b < 1 - r$). In the first case the flux from the star is unobscured. In the second case we integrate from $\kappa$ to $-\kappa$ along the boundary of the transiting body, where $\kappa$ is defined in Equation \eqref{eqn:kappa_m} for the moon and Equation \eqref{eqn:kappa_p} for the planet. To this we add the integral along the limb of the star from $\kappa^*-\pi$ to $\pi-\kappa^*$ where $\kappa^*$ is defined alongside $\kappa$ in Equations \eqref{eqn:kappa_m} and \eqref{eqn:kappa_p}. For each body we can write an expression for the obscured flux:
            \begin{eqnarray}
                \label{eqn:F_obs}
                F_\mathrm{obscured} &=& \sum_{n=0}^N u_n[G_n(\kappa, r, b) + G_n(-\kappa, r, b) \\ \nonumber &+& G_n(\kappa^*-\pi, 1, 0) + G_n(\pi - \kappa^*, 1, 0)]\\
                \nonumber &=& 2\sum_{n=0}^2 u_n\left[G_n(\pi-\kappa^*, 1, 0) - G_n(\kappa, r, b)\right]
            \end{eqnarray}
            where we have used the property that $G_n(-\phi, r, b) = G_n(\phi, r, b)$ to simplify the expression. The expression for the flux received from the star can now be given as
            \begin{equation}
                \label{eqn:no_overlap_flux}
                F = F_0 - F_\mathrm{obscured}
            \end{equation}
            where $F_0$ is the unobscured stellar flux and is given by Equation \eqref{eqn:F_0_quad}. This solution for non-overlapping bodies is equivalent to the solution given in \cite{Agol2020}.
            
        \subsection{Planet-moon overlap}
            When the planet and moon overlap each other there are a number of possible geometries that can occur, each of which require us to compute integrals along a different set of arcs. Figure \ref{fig:flowchart} is a flowchart demonstrating how we identify the correct geometry given the radii $r_p$ and $r_m$, radial distances $b_p$ and $b_m$ and the planet-moon distance $b_{pm}$. The outcomes in the flowchart, represented by the white boxes, indicate the corresponding system configuration shown in Figure \ref{fig:configurations}. Table \ref{tbl:configurations} gives the formula for the flux in each configuration. In Figure \ref{fig:flowchart} and Table \ref{tbl:configurations} the quantities $\delta_1$ and $\delta_2$ represent the distance from the center of the star to each of the planet-moon intersections. These are given by 
            \begin{eqnarray}
                \nonumber \delta_1^2 &=& r_m^2 + b_m^2 - 2r_mb_m\cos\phi_m^+,\\
                \delta_2^2 &=& r_m^2 + b_m^2 - 2r_mb_m\cos\phi_m^-.
            \end{eqnarray} where the angles $\phi_m^\pm$ are defined in Equation \eqref{eqn:phim}. The angle $\theta_{pm}$ is the angle between the line connecting the center of the star to the planet, and the line connecting the center of the star to the moon. It is computed as
            \begin{equation}
                \theta_{pm} = \mathrm{Atan2}(\Delta, (b_m - b_{pm})(b_m + b_{pm}) + b_p^2),
            \end{equation} 
            where $\Delta$ is computed from Equation \eqref{eqn:heron} with $a$, $b$, $c$ equal to $b_p$, $b_m$, and $b_{pm}$ in order from highest to lowest. The derivatives of this angle are 
            
            \begin{equation}
                \label{eqn:dthetapm_dbp}
                \frac{\partial\theta_{pm}}{\partial b_p} = -\frac{b_{pm}\sin\theta}{b_m^2},
            \end{equation}
            \begin{equation}
                \label{eqn:dthetapm_dbpm}
                \frac{\partial\theta_{pm}}{\partial b_{pm}} = (b_p\cos\theta - b_{pm})\frac{b_{pm}}{b_m^2},
            \end{equation}
            \begin{equation}
                \label{eqn:dthetapm_dtheta}
                \frac{\partial\theta_{pm}}{\partial \theta} = \frac{b_p\sin\theta}{b_m^2}.
            \end{equation}
            The factor of $b_m^2$ in the denominator of these expressions may cause concern when $b_m$ approaches zero. This singularity occurs because the angle $\theta_{pm}$ changes by $\pi/2$ instantaneously when the moon crosses the center of the star. We have no reason to expect any of the derivatives of the flux to have a discontinuity here, however, which points to the fact that this singularity must not exist in the derivatives of the flux itself, but is rather an artifact of our specific formulation. In appendix \ref{appendix:small_bm} we demonstrate that in the limit $b_m\to0$ two factors of $b_m$ arise in the numerator of the $\phi$-derivatives of the flux which cancel out with the factors of $b_m$ in the denominator such that the limit of the derivative as $b_m\to0$ is well-defined.
            
        \subsection{Normalization}
        
            In most cases we are interested in computing the normalized light curve where the unobscured flux is set to 1 and the light curve is interpreted as a measure of the fraction of the total stellar flux observed as a function of time. Equation \eqref{eqn:no_overlap_flux} and the entries in table \ref{tbl:configurations} include $F_0$, which we define as the unnormalized stellar flux. To find the normalized light curve we simply divide out $F_0$, which can be found by integrating the primitive integrals along the boundary of the star from $-\pi$ to $\pi$. This integral evaluates to 
            \begin{equation}
                \label{eqn:F_0_quad}
                F_0 = u_0\pi + u_1\frac{2\pi}{3} + u_2\frac{\pi}{2}.
            \end{equation}
            for both the quadratic and polynomial limb-darkening laws.
            
            \begin{deluxetable*}{cc}
                \tablecaption{Expressions for the total flux in each of 16 geometric cases corresponding to the diagrams in Figure \ref{fig:configurations}. To use this table, first consult Figure \ref{fig:flowchart} to determine the correct case, then look up the flux for that case. The expressions in this table are unnormalized, but can be normalized to an out-of-transit flux of $1$ by dividing out $F_0$ which is given in Equation \eqref{eqn:F_0_quad}.}
	            \tablenum{1}
                \label{tbl:configurations}
                \tablehead{\colhead{Case Label} & \colhead{Flux}}
                \startdata
                    \\
                    A & $F_0$ \\ \\
                    B & $2\sum_{n=0}^N u_n[G_n(\pi - \kappa_m^*, 1, 0) + G_n(\kappa_m, r_m, b_m)]$\\ \\
                    C & $2\sum_{n=0}^N u_n[ G_n(\pi - \kappa_p^*, 1, 0) + G_n(\kappa_p, r_p, b_p)]$\\ \\
                    D & $F_0 - 2\sum_{n=0}^N u_nG_n(\pi, r_p, b_p)$ \\ \\
                    E & $F_0 - \sum_{n=0}^N u_n[(G_n(\phi_p^-, r_p, b_p) + G_n(\phi_p^+, r_p, b_p)) - (G_n(\phi_m^-, r_m, b_m) - G_n(\phi_m^+, r_m, b_m))]$ \\ \\
                    F & $\begin{array}{lcl} \sum_{n=0}^N u_n[2G_n(\kappa_m^*, 1, 0) &-& (G_n(\phi_m^-, r_m, b_m) - G_n(-\kappa_m, r_m, b_m)) \\ &-& (G_n(\kappa_m, r_m, b_m) - G_n(\phi_m^+, r_m, b_m)) \\ &-& (G_n(\phi_p^-, r_p, b_p) - G_n(\phi_p^+, r_p, b_p))] \end{array}$\\ \\
                    G & $2\sum_{n=0}^N u_n[G_n(\pi - \kappa_p^*, 1, 0) + G_n(\kappa_p, r_p, b_p)]$\\ \\
                    H & $\begin{array}{lcl} \sum_{n=0}^N u_n[2G_n(\pi - \kappa_p^*, 1, 0) &-& (G_n(\phi_p^-, r_p, b_p) - G_n(-\kappa_p, r_p, b_p)) \\ &-& (G_n(\kappa_p, r_p, b_p) - G_n(\phi_p^+, r_p, b_p)) \\ &-& (G_n(\phi_m^-, r_m, b_m) - G_n(\phi_m^+, r_m, b_m))] \end{array}$\\ \\
                    I & $2\sum_{n=0}^N u_n[G_n(\pi - \kappa_p^*, 1, 0) + G_n(\kappa_p, r_p, b_p)]$\\ \\
                    J & $2\sum_{n=0}^N u_n[G_n(\pi - \kappa_p^*, 1, 0) + G_n(\kappa_p, r_p, b_p)]$\\ \\
                    K & $\begin{array}{lcl} \sum_{n=0}^N u_n[2G_n(\kappa_p^*, 1, 0) &-& (G_n(\phi_p^-, r_p, b_p) - G_n(-\kappa_p, r_p, b_p)) \\ &-& (G_n(\kappa_p, r_p, b_p) - G_n(\phi_p^+, r_p, b_p)) \\ &-& (G_n(\phi_m^-, r_m, b_m) - G_n(\phi_m^+, r_m, b_m))] \end{array}$\\ \\
                    L & $\begin{array}{lcl} \sum_{n=0}^N u_n[2G_n(\kappa_m^*, 1, 0) &-& (G_n(\phi_m^-, r_m, b_m) - G_n(-\kappa_m, r_m, b_m)) \\ &-& (G_n(\kappa_m, r_m, b_m) - G_n(\phi_m^+, r_m, b_m)) \\ &-& (G_n(\phi_p^-, r_p, b_p) - G_n(\phi_p^+, r_p, b_p))] \end{array} $\\ \\
                    M & $2\sum_{n=0}^N u_n[G_n(\pi - \kappa_m^*, 1, 0) + G_n(\kappa_m, r_m, b_m)]$ \\ \\
                    N & $2\sum_{n=0}^N u_n[G_n(\pi - \kappa_p^* - \kappa_m^*, 1, 0) - G_n(\kappa_p, r_p, b_p) - G_n(\kappa_m, r_m, b_m)]$ \\ \\
                    O & $\begin{array}{lcl} \sum_{n=0}^N u_n[2G_n(\pi - \kappa_p^* - \kappa_m^*, 1, 0) &-& (G_n(\phi_m^+, r_m, b_m) - G_n(-\kappa_m, r_m, b_m)) \\ &-& (G_n(\kappa_m, r_m, b_m) - G_n(-\phi_m^-, r_m, b_m)) \\ &-& (G_n(\kappa_p, r_p, b_p) - G_n(\phi_p^+, r_p, b_p)) \\ &-& (G_n(\phi_p^-, r_p, b_p) - G_n(-\kappa_p, r_p, b_p))]\end{array}$\\ \\
                    P & $\begin{array}{lcl} \sum_{n=0}^N u_n[2G_n(\pi - (\kappa_p^* + \kappa_m^* + \theta_{pm})/2, 1, 0) &-& (G_n(\kappa_m, r_m, b_m) - G_n(-\phi_m^-, r_m, b_m)) \\ &-& (G_n(\phi_p^-, r_p, b_p) - G_n(-\kappa_p, r_b, b_p))]\end{array}$\\ \\
                \enddata
            \end{deluxetable*}
            
    \section{Implementation Details}
        \subsection{Keplerian Dynamics}
            \label{sec:kepler}
        
         We implement two separate dynamics models which apply to different physical systems in which mutual transits might be observed. The first is a hierarchical model in which a primary body is orbited by a secondary body which together orbit the central mass. This model is applicable to exomoons, binary planets, and hierarchical triple-star systems. The second is a confocal model in which two bodies independently orbit a central mass, which is applicable to multi-planet systems in which multiple planets transit the host star. We describe each of these models in detail below. For both models we solve Kepler's equation using the solver from \cite{ForemanMackey2021}, which is itself based on the Kepler solver described in \cite{RaposoPulido2017}. We propagate derivatives of the mean and eccentric anomalies with respect to each of the input parameters through the formula for the Cartesian coordinates of each body using the chain rule. For both types of system our dynamics module outputs $b_p$, $b_{pm}$, $\theta$, and the derivatives of each with respect to all input parameters. These coordinates are then used as input to the photometry module. The derivatives of the flux with respect to each of the input parameters are then given by the chain rule as 
            \begin{equation}
                \frac{dF}{dq_i} = \frac{\partial F}{\partial b_p}\frac{db_p}{dq_i} + \frac{\partial F}{\partial b_{pm}}\frac{db_{pm}}{dq_i} + \frac{\partial F}{\partial \theta}\frac{d\theta}{dq_i}
            \end{equation}
            where $q_i$ is the $i^\mathrm{th}$ input parameter to the Kepler module. 
        
        \subsubsection{Hierarchical Systems}
        
            For hierarchical systems we adopt a simplified model of the dynamics in which we assume the system consists only of the star, planet, and moon. We further assume no interaction between the moon and the star. This simplifies the computation significantly because it allows us to model two Keplerian systems, one consisting of the star and the planet/moon combined mass and the other consisting of the planet and moon separately, rather than requiring us to simulate the three-body dynamics of the system.
            
            The approximation of the system as a hierarchy of two-body Keplerian systems is valid in the case that the moon orbits well within the planet's Hill sphere. Because a moon orbiting too close to the outer edge of the Hill sphere is unstable over long timescales we would expect that any exomoon we observe would meet this criteria. Our approximation excludes non-Keplerian forces that occur in three-body systems, so care should be taken in cases where the precession rate is expected to be high relative to the duration of observations. The nodal precession period due to three-body moon-planet-star interactions is given by \cite{Martin2019} as
            \begin{equation}
                P_\mathrm{precess} = \frac{4}{3}\frac{M_p + M_*}{M_*}\frac{P_p^2}{P_m}\frac{1}{\cos i_m}.
            \end{equation}
            If we assume that the mass of the planet is negligible compared to that of the star, then we can use the following criterion to determine when precession is not expected to impact our ability to model the system:
            \begin{equation}
                \frac{P_\mathrm{precess}}{P_p}\approx \frac{4}{3}\frac{P_p}{P_m}\frac{1}{\cos i_m} \gg 1.
            \end{equation}
            When this criterion is not met we may observe nodal precession on the timescale of the planetary period, causing the parameters of the moon's orbit to change from transit to transit and thus invalidating our dynamical model. In this case the user may wish to provide a dynamics module of their own which takes into account three-body interactions.
            
        \subsubsection{Confocal Systems}
        
            For confocal systems we assume no interactions between the two bodies and solve Kepler's equation for each body separately. For the first body we set the longitude of the ascending node, $\Omega$, equal to $180$ degrees following \cite{Winn2010}. For the second body we allow $\Omega$ to vary since this gives us the mutual inclination between the two orbits.
            
            Because we assume no interaction our code is not able to model transit timing variations. However, since our primary purpose is to model mutual transits and the timing variation due to the other body is small when the bodies transit simultaneously, this should not be a problem in most cases. One scenario in which this might present an issue is when other planets are present in the system and contribute to transit timing variability.
            
    \subsection{Photometric Model}
        
        Given the positions of each of the two occulting bodies at a point in time we use the decision tree in Figure \ref{fig:flowchart} to identify the appropriate geometric case. This determines the formula for the flux integral which can be looked up in Table \ref{tbl:configurations}. 
        
        The linear and polynomial terms of the flux integral include incomplete elliptic integrals. Instead of computing these integrals directly, which can result in numerical instabilities for some values of the inputs, we compute them in terms of the Bulirsch elliptic integrals $el1$, $el2$, and $el3$, which are related to the standard elliptic integrals $F$, $E$, and $\Pi$ as follows:
        \begin{eqnarray}
            \nonumber F(\phi, k) &=& el1(x, k_c),\\
            E(\phi, k) &=& el2(x, k_c, 1, k_c^2),\\
            \nonumber \Pi(\phi, n, k) &=& el3(x, k_c, p),
        \end{eqnarray}
        where $x = \arctan\phi$, $k_c = \sqrt{1-k^2}$, and $p = n + 1$ \citep{Bulirsch1965el1el2, Bulirsch1969el3}. When derivatives are not required we use the following relationship to compute a linear combination of $F(\phi, k)$ and $E(\phi, k)$ with only one evaluation of $el2$ \citep{Bulirsch1969lincombo}:
        \begin{equation}
            \alpha E(\phi, k) + \beta F(\phi, k) = el2(x, k_c, \alpha + \beta, \beta + \alpha k_c^2).
        \end{equation}
        When computing the derivatives of the model all three of the incomplete elliptic integrals are required in different linear combinations. This requires us to compute all three functions individually, but the integrals can be computed simultaneously for computational efficiency by using some components of the computation which are shared in common.
        
    \subsection{Exposure time integration}
        
        Telescope observations do not represent instantaneous snapshots of a system, but are instead integrated over the duration of the exposure. Failing to account for finite integration times can negatively affect our ability to measure parameters of the system \citep{Kippin2010}. To capture this effect, we include in our model an option to numerically integrate the light curve using either the trapezoidal rule or Simpson's rule. 
        
\begin{figure*}
            \centering
            \includegraphics[width=\textwidth]{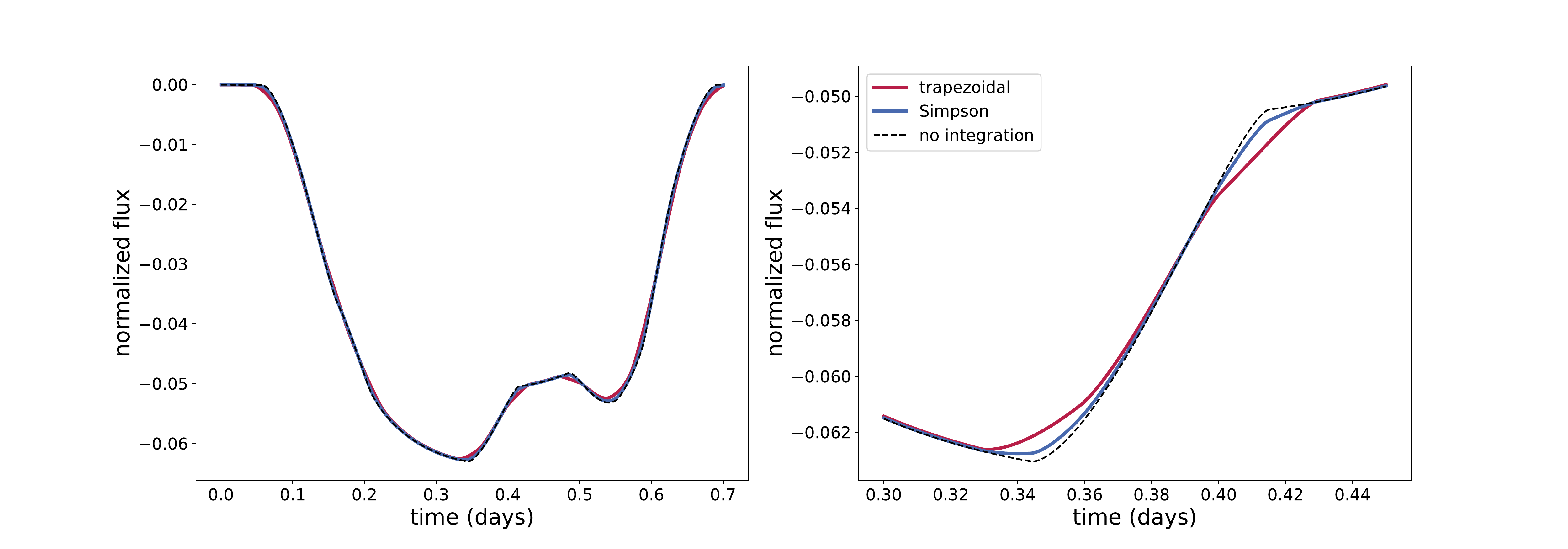}
            \caption{An exposure-time integrated light curve. The left panel shows the full light curve with the dashed line being the un-integrated curve. The red line is the integrated curve produced by the trapezoidal rule and the blue line is the integrated curve produced by Simpson's rule. On the right we have zoomed in to the part of the transit where the moon crosses the limb of the planet. At this point the flux changes sharply which produces errors in the time-integration, which accounts for the difference between the trapezoidal curve and the Simpson's curve. Simpson's method is a higher order method and is more accurate, especially at these contact points.} 
            \label{fig:exp_time_int}
        \end{figure*}

        For the trapezoidal rule with arbitrary exposure times we up-sample the light curve by a factor of two and compute the observed flux at a time $t$ as
        \begin{equation}
            F_\mathrm{obs}(t) = \frac{1}{2}\left[F\left(t - \frac{\Delta t}{2}\right) + F\left(t + \frac{\Delta t}{2}\right)\right],
        \end{equation}
        where $\Delta t$ is the exposure time. This computation comes at the cost of twice the number of evaluations of the flux. However, for the common case where the exposure time is equal to the time between observations the start times and end times for adjacent exposures are the same. In this case we need only compute $N + 2$ flux evaluations where $N$ is the number of input times and the $2$ arises from the need to evaluate the flux at the beginning of the first integration and the end of the last integration. If derivatives are required these can be integrated in the same way as the flux. Because the trapezoidal rule computes the integral of a linear spline interpolation of the true light curve the result can be inaccurate at points where the derivative of the light curve with respect to time changes rapidly. This occurs at the contact points between the star and planet, star and moon, and planet and moon. 
        A more accurate option is Simpson's rule. For Simpson's rule in the case of arbitrary exposure times we up-sample the light curve by a factor of three and compute the observed flux as 
        \begin{equation}
            F_\mathrm{obs}(t) = \frac{1}{6}\left[F\left(t-\frac{\Delta t}{2}\right) + 4F\left(t\right) + F\left(t+\frac{\Delta t}{2}\right)\right].
        \end{equation} 
        This costs three times the number of flux evaluations as the un-integrated light curve, and $50\%$ more than for the trapezoidal rule, but it will be much more accurate than the trapezoidal rule for long integration times. Figure \ref{fig:exp_time_int} shows the effect of exposure time integration for an exposure time of 0.03 days on a light curve identical to the light curve in Figure \ref{fig:demo}. The left panel of the figure shows the full light curve while the right panel zooms in on the point where the moon transits the planet to show the effect of the sharp transition in the flux on the integral produced by the two methods.
        
    \section{Speed}
        \label{sec:speed}
        
        We have benchmarked \texttt{gefera} on an Intel Macbook Pro from early 2015 with a 3.1 GHz Dual-Core Intel Core i7 CPU. Because the number of integrals needed to compute the flux depends on the geometry of the planet and moon, the time to compute the model is dependent on the geometry of the transit. We have therefore benchmarked two models, a simple model and a complex model. The simple model simulates an event in which the planet and moon do not overlap each other at any point during the transit, whereas the complex model simulates a transit in which the moon and planet overlap while crossing the limb of the star. This is the most computationally expensive case, in which the full set of three incomplete elliptic integrals must be computed for as many as four separate arcs at some timesteps. In comparison the simple model will only require two complete elliptic integrals to be computed for each timestep during which the moon and planet both overlap the star. 
        
\begin{figure*}
            \centering
            \includegraphics[width=\textwidth]{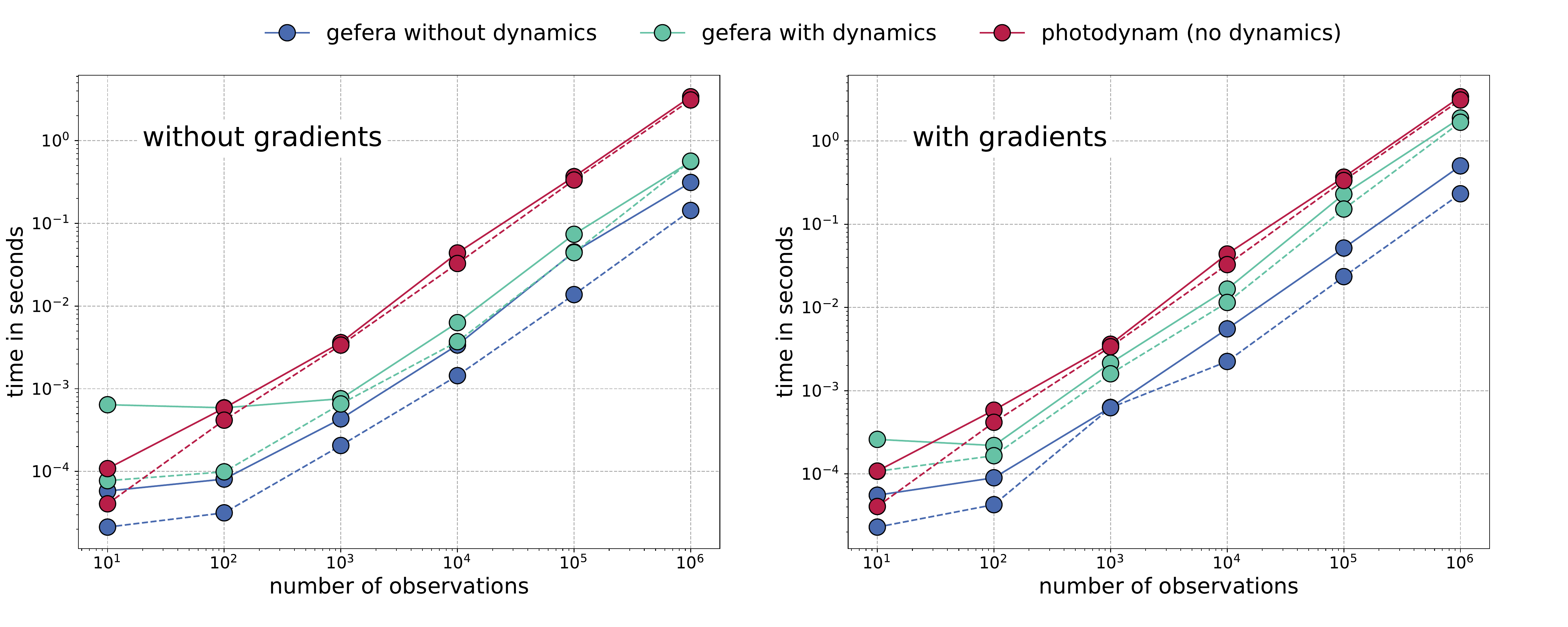}
            \caption{Benchmarks for \texttt{gefera} and \texttt{photodynam}. The right panel includes the gradient calculation whereas the left panel does not. Since \texttt{photodynam} does not include gradients these benchmarks are the same for both panels. Dashed lines indicate that a ``simple'' transit was simulated in which the star and planet do not overlap at any point during the transit. Solid lines indicate a ``complex'' model where the moon and planet overlap during their egress. For \texttt{gefera} we include benchmarks both with and without the inclusion of the dynamics module, whereas for \texttt{photodynam} we include benchmarks only for the photometry module.} 
            \label{fig:benchmarks}
        \end{figure*}

        Figure \ref{fig:benchmarks} shows our benchmarks for a number of observations ranging from ten to one-million for both the simple (dashed lines) and complex (solid lines) models. In the left panel we compute our model without gradients, and in the right we compute the model with gradients. We also include benchmarks for the \texttt{photodynam} code \citep{pal2012, Carter2011} which implements a similar method to our code but is written to be more general and can include any number of bodies (as opposed to \texttt{gefera} which only allows for two bodies in addition to the star). While \texttt{photodynam} does not include gradients, we have plotted the benchmarks in both panels for the sake of comparison. We find that \texttt{gefera} is about 22 times faster than \texttt{photodynam} without computing gradients for the simple model, and 10 times faster for the complex model without gradients. When we include the gradient computation, \texttt{gefera} is about 15 times faster than \texttt{photodynam} for the simple model and 7 times faster for the complex model. These results hold for all runs with more than about 100-1,000 timesteps. For very small numbers of timesteps \texttt{photodynam} performs similarly to, but still slightly slower than, \texttt{gefera}.
        
        We benchmark \texttt{gefera} with and without the inclusion of the dynamics module. While the \texttt{photodynam} code does include dynamics, we have chosen to time only the photometry module. This is because the dynamics module provided with \texttt{photodynam} is not well-suited to reproducing the hierarchical sun-planet-moon system that we implement in \texttt{gefera} and as a result a comparison between the two would not be very informative. The benchmarks for \texttt{photodynam} should therefore be compared to the results for \texttt{gefera} without dynamics (the dark blue lines) rather than with dynamics. 
        
        We were unable to include benchmarks for \texttt{LUNA} as we do not have access to the fully optimized \texttt{LUNA} code. Our implementation of the \texttt{LUNA} algorithm was written in order to test accuracy, not speed, and is likely not as efficient as the original implementation. Therefore it does not seem fair or useful to provide a comparison. 
            
    \section{Comparison to \texttt{LUNA}}
        \label{sec:luna}
            \texttt{LUNA} \citep{Kipping2011} is an algorithm for computing exoplanet/moon transits which applies the small-planet approximation \citep{Mandel2002} for the flux blocked by the area of the moon's overlap with the star while using the full analytic transit solution for the area of the planet/star overlap. While the original implementation of \texttt{LUNA} has not been made publicly available, we have implemented our own version of the algorithm as published in \cite{Kipping2011} for purposes of comparison. 
            
\begin{figure*}
                \centering
                \includegraphics[width=\textwidth]{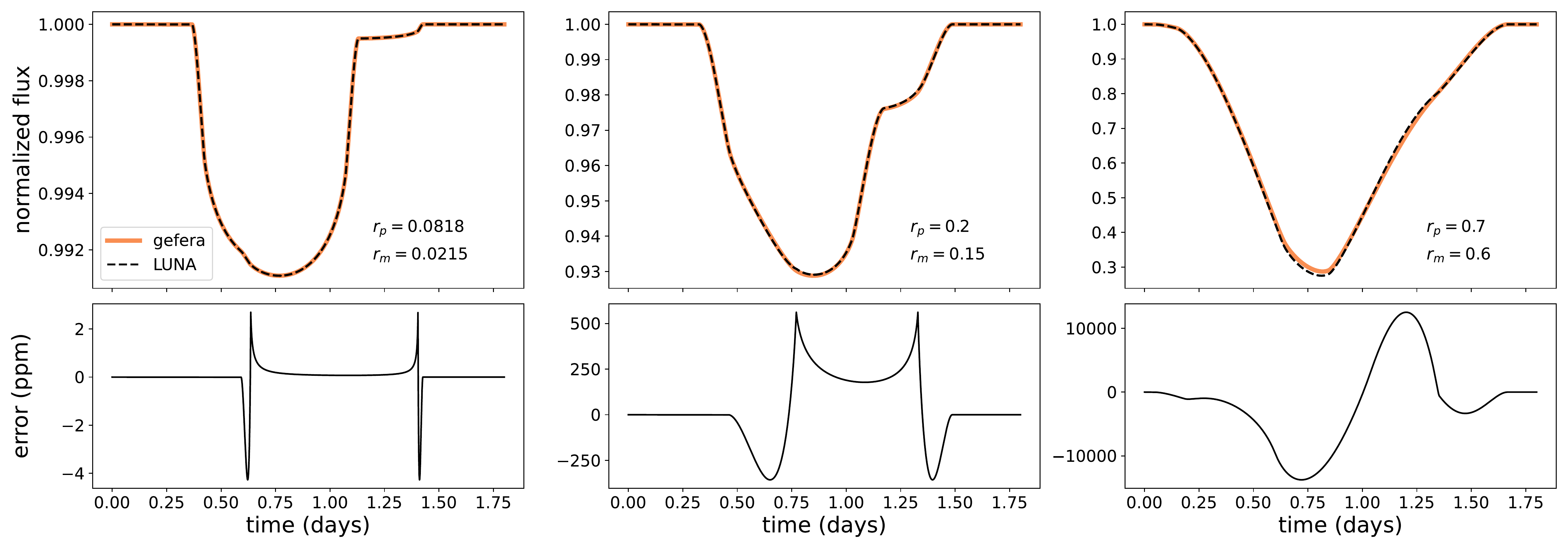}
                \caption{\texttt{LUNA} and \texttt{gefera} light curves and error in ppm, labeled by the radii of the planet and moon in units of the stellar radius. \textbf{Left:} Radii for the moon and planet are taken from the maximum likelihood solution for Kepler-1708 b i given in \citet{Kipping2022}. In this case the difference between the \texttt{LUNA} solution and the \texttt{gefera} solution is only about four parts-per-million at most. \textbf{Center:} The same transit with a much larger planet and moon. Here the maximum difference between the two solutions is 500 parts-per-million. \textbf{Right:} The same transit again with a ``planet'' and ``moon'' approaching the radius of the star. These parameters may be reasonable for an eclipsing triple-star system. In this case the small-planet approximation breaks down with the error approaching 1\%.} 
                \label{fig:LUNA_comparison_1708b}
            \end{figure*}

            As a test, we model the transit of an exoplanet and accompanying moon with parameters consistent with the parameters inferred for the exomoon candidate Kepler-1701b i as published in \cite{Kipping2022}. We adopt the mean of the posterior distribution for the planet-star radius ratio, $0.0818$ and the moon-planet radius ratio $0.263$. We find a maximum discrepancy of about 4 ppm between \texttt{LUNA} and our model, which demonstrates the applicability of the small-moon approximation for this case. The first panel of Figure \ref{fig:LUNA_comparison_1708b} shows the \texttt{LUNA} and \texttt{gefera} light curves with the absolute error between the two models. In the second panel we increase the radii of the planet and moon to 0.2 and 0.15 $R_*$ respectively. Here the small-moon approximation begins to break down with the \texttt{LUNA} model differing by about 500 ppm from the \texttt{gefera} model. In the third panel we again increase the radii to 0.7 and 0.6 $R_*$ respectively. While these parameters are unlikely to occur for a realistic binary planet system, it serves to demonstrate the breakdown of the small-planet approximation for large bodies. 
            
            We conclude that \texttt{LUNA} is likely sufficiently accurate for the vast majority of realistic planet/moon systems, but the difference between the approximate light curve computed by \texttt{LUNA} and the exact light curve computed by \texttt{gefera} may become important for large moons or binary planet systems, or in the case that very high precision photometry is available. 
            
\begin{figure}
                \centering
                \includegraphics[width=\hsize]{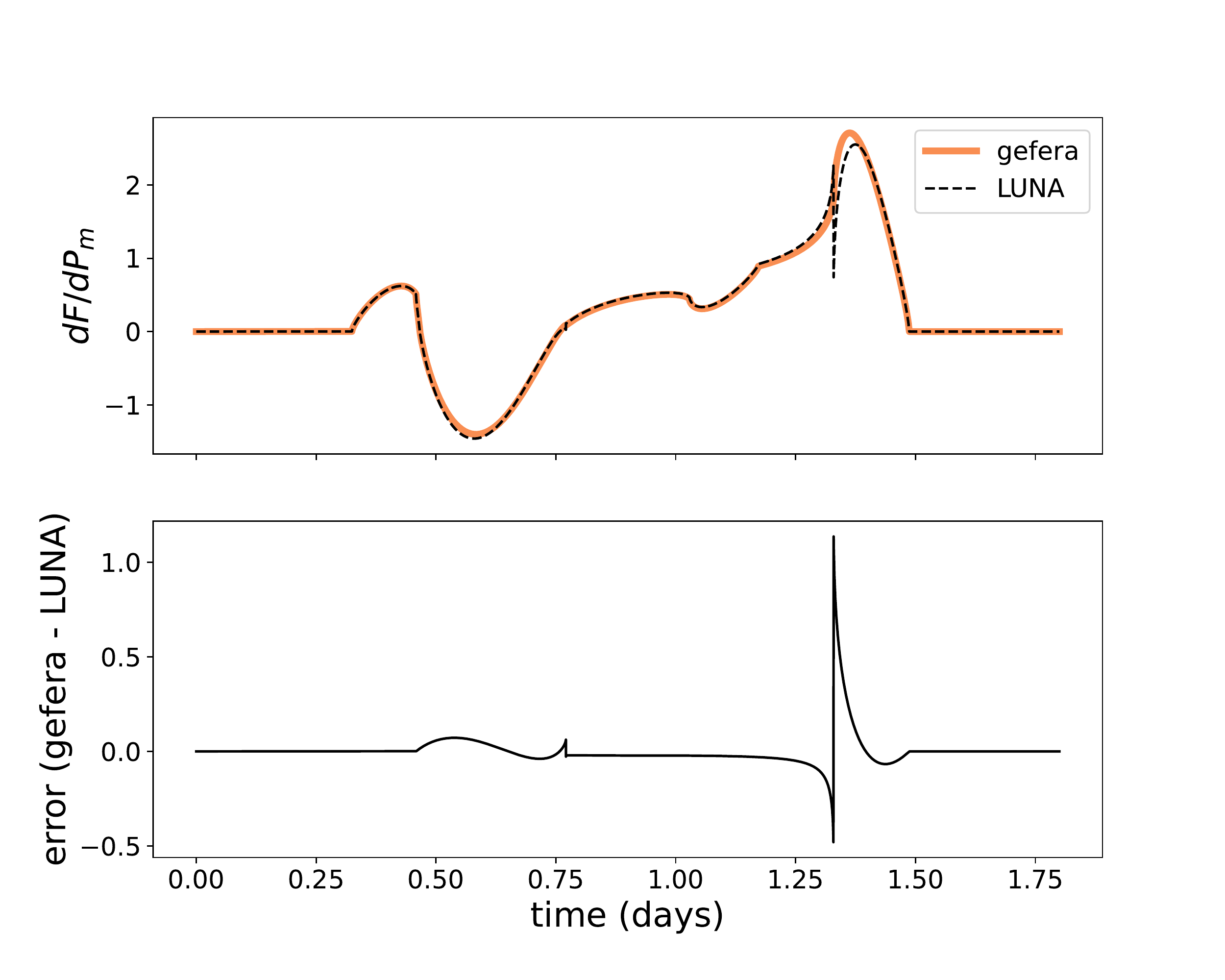}
                \caption{\texttt{LUNA} and \texttt{gefera} derivatives and the difference between them. The orbital parameters are identical to those used in Figure \ref{fig:LUNA_comparison_1708b} and the radii of the planet and moon are $0.2R_*$ and $0.15R_*$ respectively, matching the center panel of Figure \ref{fig:LUNA_comparison_1708b}. The error in the derivative is larger than for the light curve itself, crossing the one percent threshold at the point where the moon contacts the inner edge of the stellar limb.} 
                \label{fig:LUNA_derivative}
            \end{figure}

\begin{figure}
                \centering
                \includegraphics[width=\hsize]{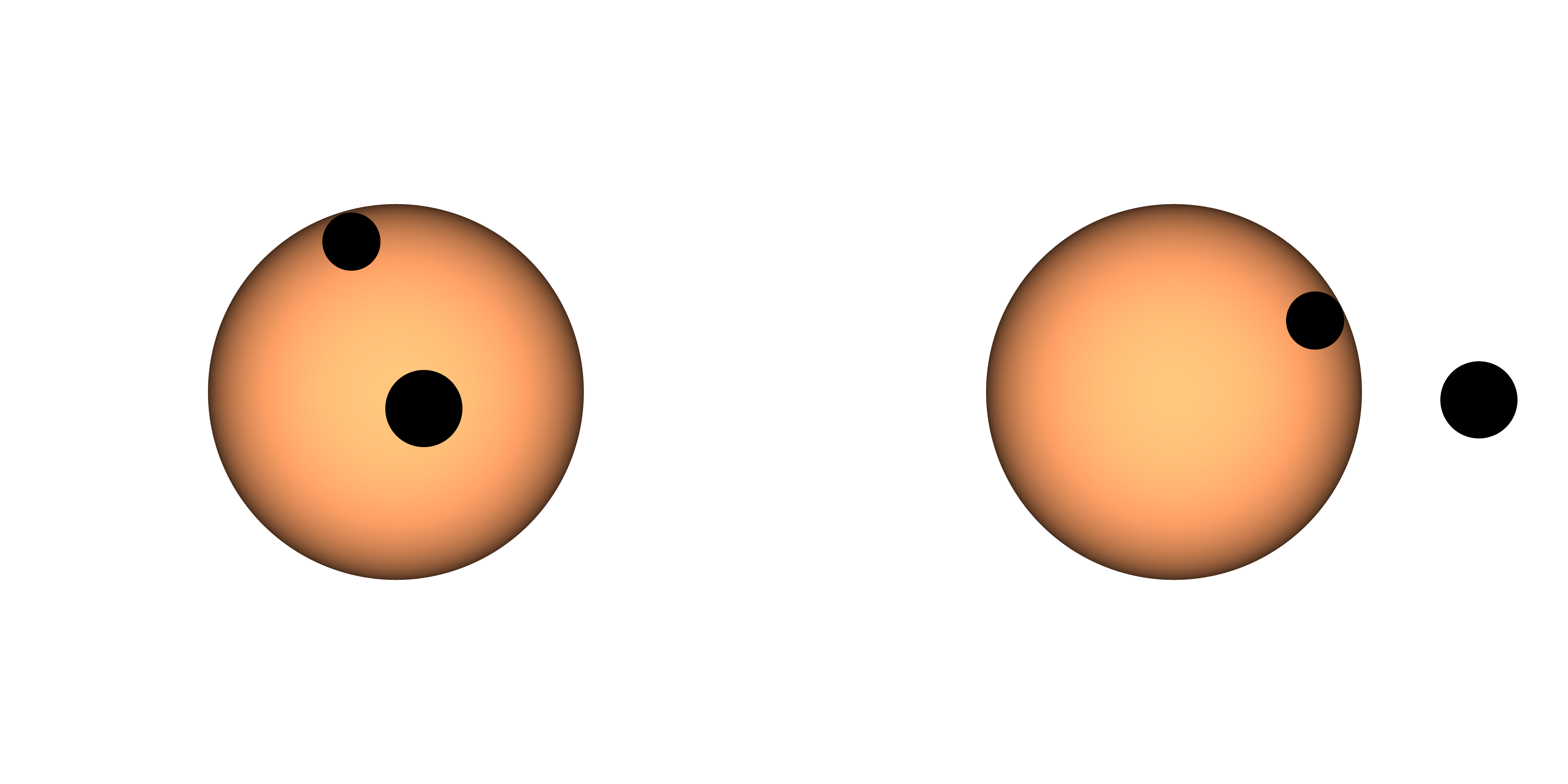}
                \caption{System configuration at the points where the derivative diverges in Figure \ref{fig:LUNA_derivative}. The left panel shows the first point of divergence and the right the second.} 
                \label{fig:LUNA_contact}
            \end{figure}

            Another outcome of the use of the small-moon approximation in \texttt{LUNA} is that a \texttt{LUNA} light curve does not have a continuous derivative with respect to the position or radius of the moon. The derivatives of the \texttt{LUNA} model diverge at the contact points between the moon and the inner edge of the stellar limb. To illustrate this we compute the derivative of the \texttt{LUNA} and \texttt{gefera} models with respect to the period of the moon's orbit using finite differences using the parameters from the central panel of Figure \ref{fig:LUNA_comparison_1708b}. The upper panel of Figure \ref{fig:LUNA_derivative} shows the derivative itself while the lower panel shows the difference between the \texttt{LUNA} and gefera derivatives. The configuration of the system at the points of divergence are illustrated in Figure \ref{fig:LUNA_contact}. The lack of continuous derivatives may create issues for gradient-based inference algorithms such as Hamiltonian MCMC and any other analysis that requires the use of model derivatives.
            
            We are unable to benchmark the computational speed of \texttt{LUNA} since we don't have access to the fully optimized version of the code, but we note that \texttt{gefera} is fast enough to enable MCMC inference for most datasets on a laptop computer, so we argue that any reduction in computational speed between \texttt{LUNA} and \texttt{gefera} is likely justified by the increase in accuracy. Additionally, since \texttt{gefera} provides derivatives of the transit it can be used with Hamiltonian MCMC which, carried out with No U-turn Sampling (NUTS), may be more efficient than  standard MCMC.
            
            One final note on \texttt{LUNA} is that in the process of implementing the algorithm we identified an error in equation 37 of \cite{Kipping2011} that appears to have been carried over from an incorrect formula published in \cite{Fewell2006} for the area of overlap between three circles in the case that the region of overlap includes more than half of the area of the smallest circle. In appendix \ref{appendix:fewell_correction} we provide an explanation of the error and a corrected formula.
            
    \section{Applications}
        \label{sec:applications}
        We envision our model being used to search for exomoons in observations from current and upcoming missions including JWST, TESS, and PLATO, as well as archival data from the Kepler and CoRoT missions. It also may be useful for analyzing simultaneous mutual transits of eclipsing binary stars and circumbinary planets or more complicated multi-star systems \citep{Carter2011}. In the next section we briefly discuss the options for conducting Bayesian inference using our model. Because we provide derivatives of the planet/moon transit, our model also may be used to conduct an information analysis along the lines of \cite{Carter2008} or \cite{Price2014}.
        
        It is also our hope that providing a publicly available exomoon transit code will make it easier for the community to conduct independent analyses of exomoon candidates such as Kepler-1625b i \citep{Teachey2018} and Kepler-1708b i \citep{Kipping2022}. These candidates raise the exciting possibility that exomoon detections lay within the grasp of existing datasets, but the lack of non-proprietary exomoon transit models has made it difficult for other groups to conduct their own searches or to follow up on existing candidates. 
            
    \subsection{Inference with \texttt{gefera}}
        \label{sec:inference}
        
        The \texttt{gefera} codebase is provided as a pip-installable python package. It is written to have a simple python interface which allows the user to define the orbit of the moon-planet system about the star and the orbit of the moon about the planet using standard keplerian orbital parameters. The user can then compute a light curve with or without derivatives. This interface is useful both for quickly computing and plotting sample light curves as well as for plugging into inference packages for MCMC or nested sampling. With access to the derivatives users may also make use of gradient-based inference algorithms such as Hamiltonian Monte-Carlo with No-U-Turn Sampling \citep[NUTS;][]{Hoffman2014}. In testing the code we found that the streamlined python interface used with \texttt{emcee} \citep{ForemanMackey2019} is the simplest and most efficient way to conduct inference for a moon/planet transit model. However, in the future we intend to implement a set of \texttt{aesara} ops \citep{Willard2021} which will allow \texttt{gefera} to be used with \texttt{pymc} \citep{Salvatier2016}. We will also explore the possibility of adding interfaces for other popular inference packages in future versions of our \texttt{gefera}. Making use of the more advanced gradient-based samplers provided by \texttt{pymc} may also be advantageous for larger models combining multiple planets or binary planet/moon systems in which the number of parameters can become very large. 
        
    \subsection{Using \texttt{Gefera} to Model Binary or Triple-star Systems}
        \label{sec:triplestars}
    
        We have briefly mentioned that \texttt{gefera} may be used to model stellar transits in binary or triple-star systems in addition to moon-planet and planet-planet transits. We now explain how \texttt{gefera} can be used to model such a transit.
        
        When one of the transiting bodies emits its own flux, we must first consider whether that body is itself being transited (i.e. is the body emitting flux positioned between the two other bodies, or is it the nearest to the observer). If the second flux-emitting body is the closest body to the observer and is not itself being transited, then there is no need for a special procedure. This type of transit can be directly modeled in \texttt{gefera} using the same prescription outlined for planet-planet or planet-moon transits. However, when the second flux-emitting body is positioned between the two other bodies in the system, as in the case of a coplanar circumbinary system, we must consider the second transit of this luminous body by the frontmost transiting body. We can do this by first computing the mutual transit under the assumption that neither transiting body is luminous. To this flux we can add the flux from a single transit between the second star and the front-most transiting body. The total flux can be expressed as 
        \begin{equation}
            F_\mathrm{total} = F_\mathrm{non-luminous} + F_\mathrm{two\ bodies}
        \end{equation}
        where $F_\mathrm{non-luminous}$ refers to the \texttt{gefera} model computed under the assumption that the two transiting bodies are non-luminous and $F_\mathrm{two\ bodies}$ refers to the ordinary two-body transit between the second star and the front-most star or planet.
        
    \subsection{Test Case: Kepler-51 b \& d}
        \label{sec:kepler51}
        
        As a test of our code we have conducted a short re-analysis of a possible mutual transit event involving Kepler-51 b \& d, previously studied by \citet[][hereafter M14]{Masuda2014}. Mutual transit events are characterized by a ``bump'' -- an increase in flux during the transit resulting from the decrease in the obstructed area of the stellar disk during the planet-planet overlap. 
        Unfortunately bumps may also result from starspot crossings making the interpretation of such a feature ambiguous. Ultimately M14 concludes that a starspot crossing is the likely cause of this feature, basing their conclusion on several arguments. First, fitting the mutual transit event, M14 infers impact parameters for both planets near zero and a large mutual inclination of $-25$ degrees between the two orbits. 
        The impact parameters conflict with the values inferred from other transits of planets b and d which are close to 0.25 for both planets, despite agreeing with the Kepler team's determination of $b=0.03$ for planet b and $b=0.061$ for planet d\footnote{Data retrieved from the NASA Exoplanet Archive \citep{planetary_systems_table}}. These large impact parameters reduce the likelihood of a mutual transit, while impact parameters close to zero increase the likelihood of the event. 
        Second, large mutual inclination is a problem for the mutual transit interpretation as such a configuration is a priori unlikely to have both planets transit and would further reduce the chance of planet-planet overlap during the simultaneous transit.
        
        M14 models the mutual transit bump by computing the area of overlap between the two planets, then multiplying that by a factor related to the surface brightness of the star at the location of the overlap as specified by a quadratic limb-darkening law, and finally adding this flux back to the light curve (since it is double-subtracted when the planet-planet interaction is not accounted for).
        This is similar to the approach taken by \cite{Kipping2011} based on the small-planet approximation of \cite{Mandel2002}, except that the approximation is applied only to the area of overlap rather than to the entirety of the smaller body. 
        
\begin{figure}
                \centering
                \includegraphics[width=\hsize]{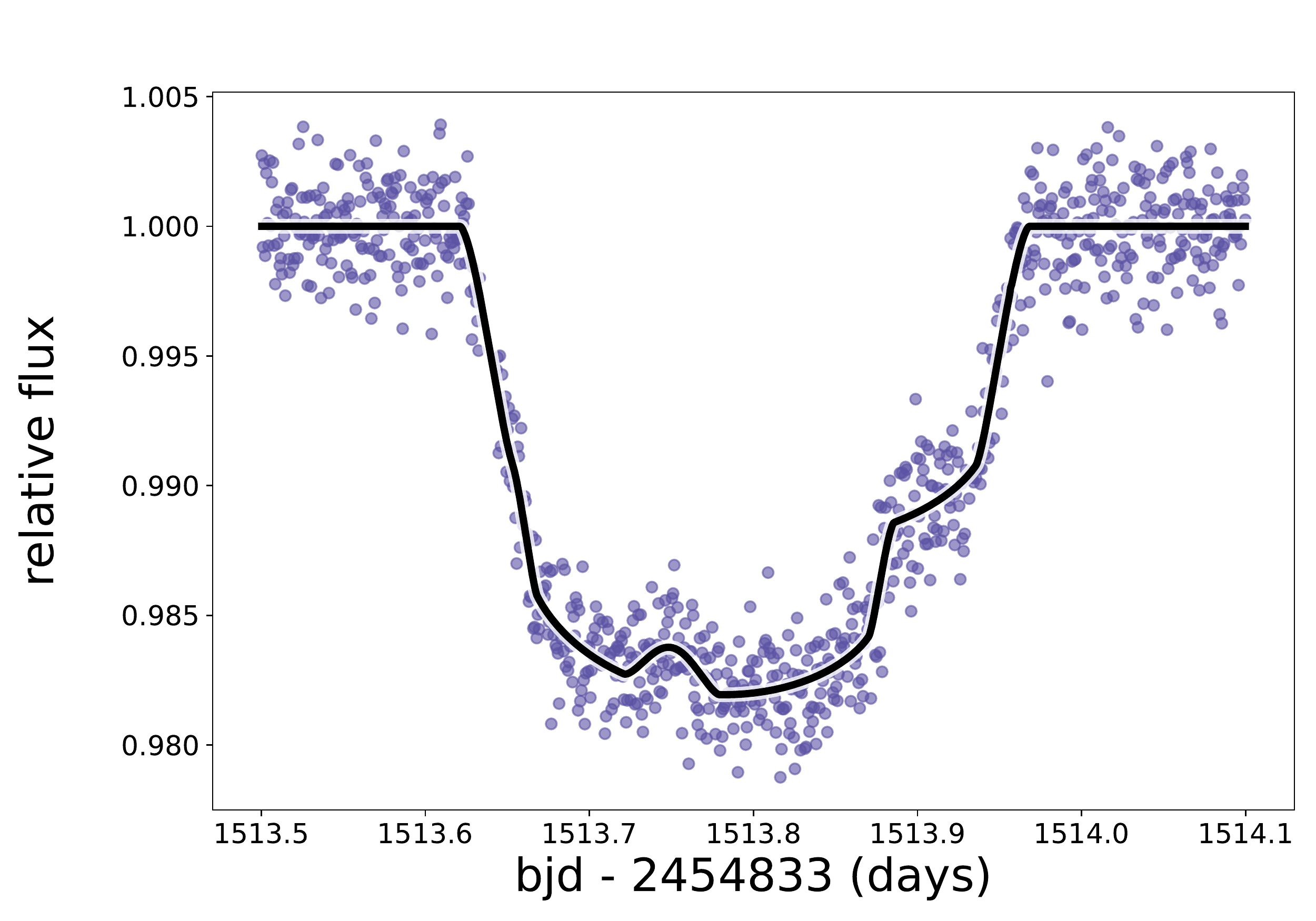}
                \caption{Best fit to the Kepler observations of the Kepler-51 system. The fit was found using the truncated Newtonian algorithm provided by \texttt{scipy} with derivatives.} 
                \label{fig:kep51_fit}
        \end{figure}

        We begin our own analysis by using the built-in methods from \texttt{lightkurve} \citep{lightkurve} to detrend the short-cadence PDCSAP Kepler flux using a Savitzky-Golay filter with a polynomial of order 2 and a window length of 1.65 hours. We mask the in-transit points prior to detrending. We then use the truncated Newtonian minimization routine provided by \texttt{scipy} \citep{scipy} to minimize a two-planet transit model with respect to the Keplerian orbital parameters $\{a, t_0, e, P, \omega, i\}$ for each body where $a$ is the semimajor amplitude, $t_0$ is the reference epoch, $e$ the eccentricity, $P$ the period, $\omega$ the longitude of periapse, and $i$ the inclination. In addition to these we include $\Omega$, which is the mutual inclination between the two orbits, as well as the radii of both bodies and the two quadratic limb-darkening parameters ($c_1$ and $c_2$ in Equation \eqref{eqn:intensity}). The truncated Newtonian minimization routine accepts the derivatives of the likelihood with respect to the parameters which we find speeds up the minimization routine by a factor of 10 or more. The best-fit model is shown in Figure \ref{fig:kep51_fit}, and Figure \ref{fig:kep51_snaps} shows the best-fit transit as a series of snapshots forming a ``movie'' of the event. These snapshots can be compared to Figure 3 in M14, with the caveat that our snapshots are rotated counterclockwise through an angle $\Omega$ relative to the figure in M14. This rotation does not affect the appearance of the light curve. 
        
\begin{figure}
                \centering
                \includegraphics[width=\hsize]{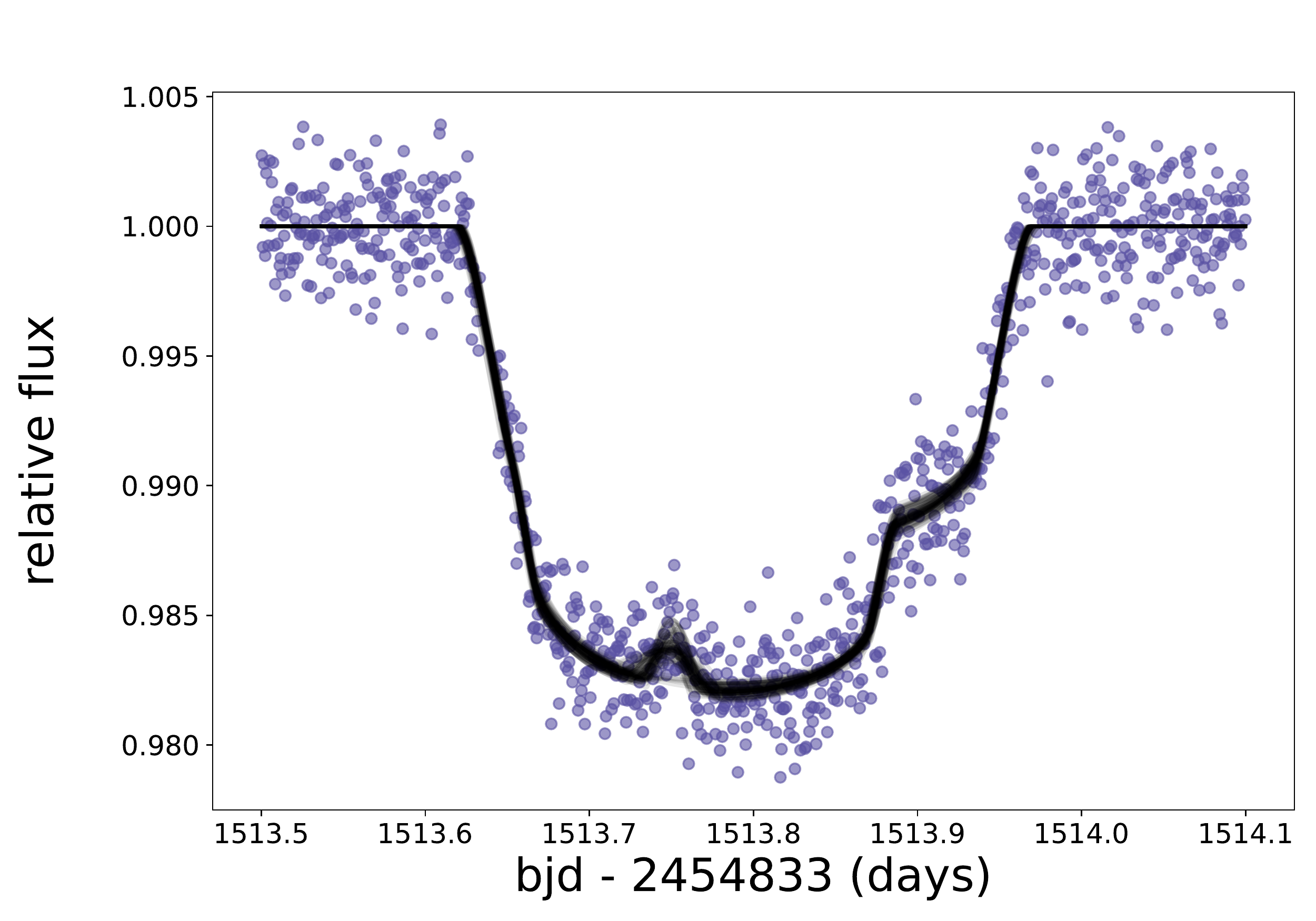}
                \caption{Kepler-51 photometric time series observed by Kepler (purple points) with 100 randomly drawn samples from the posterior distribution (black lines).} 
                \label{fig:kep51_samples}
        \end{figure}

\begin{figure*}
                \centering
                \includegraphics[width=\textwidth]{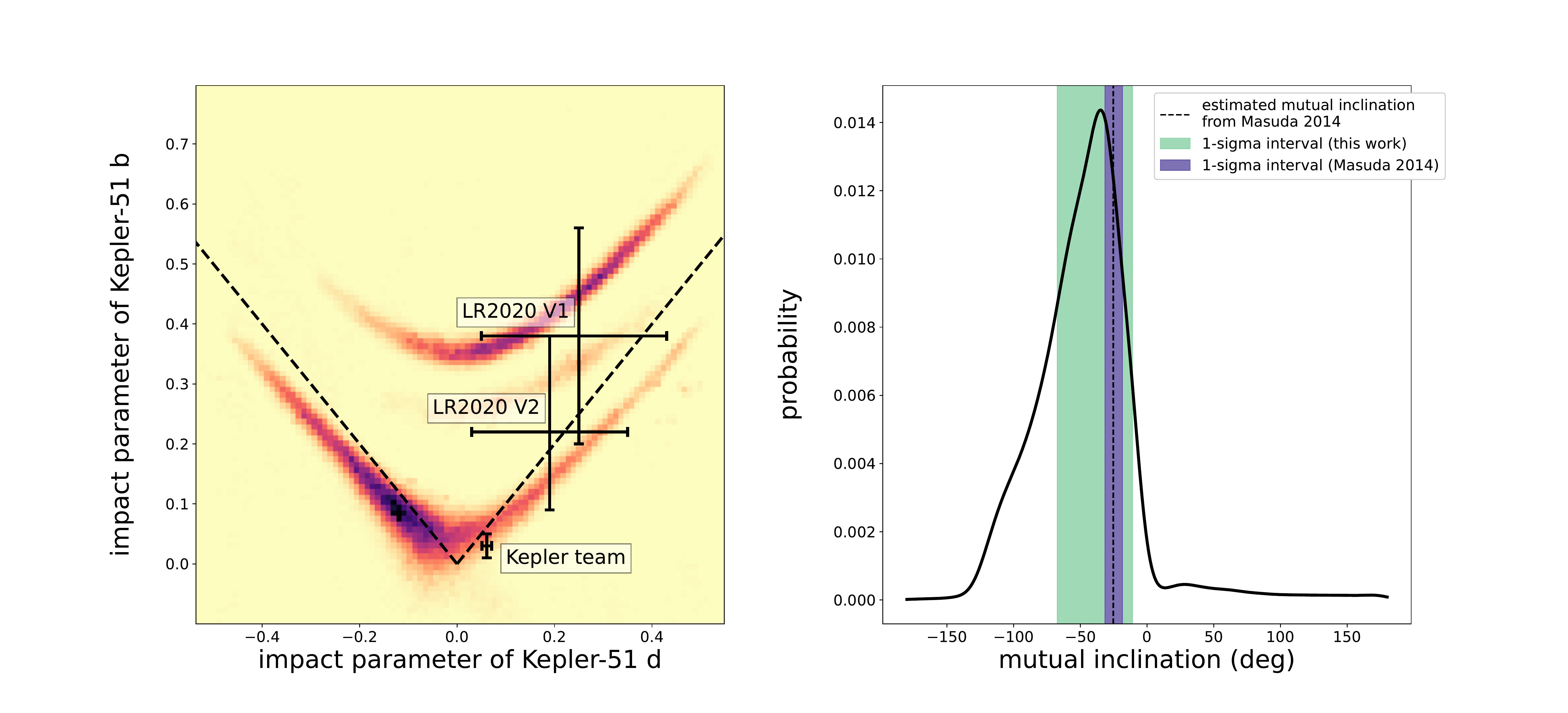}
                \caption{Posterior plots for the impact parameter of each planet versus the mutual inclination.
                \textbf{Left:} Joint posterior distribution for the impact parameters of planets b \& d with darker colors indicating higher probability. The distribution is multi-modal, which presents a difficulty for MCMC methods. Dashed lines represent the line along which the impact parameters for both planets are equal, and the error bars represent the impact parameters found by the Kepler team and for each of the visits (V1 and V2) with \textit{HST} from \citet{Libby-Roberts2020}. This figure can be compared to Figure 4 in \citet{Masuda2014}. We have made no attempt to account for the possible failure of the MCMC chains to fully explore all of the modes. Therefore the apparent densities of each mode relative to the others should not be interpreted as a preference for one mode over another. We merely intend to show that multiple modes exist and to demonstrate the use of our model with an MCMC sampler. \textbf{Upper right}: Kernel density estimate of the posterior for the mutual inclination between planets b \& d marginalized over all other parameters. The dashed line is the estimate from \cite{Masuda2014} and the purple shaded region shows the 1-sigma confidence interval from that paper. The green shaded region represents the 1-sigma confidence interval from this work. We find agreement between our determination of the mutual inclination and the value found by M14, with the exception that our posterior admits a wider range of mutual inclination angles than was found by M14. Again, we caution that the multi-modality of the posterior presents an issue for our MCMC analysis and thus the relative heights of the modes in this plot may not represent the actual relative probabilities of true transit parameters having been drawn from each mode.} 
                \label{fig:kep51_posterior}
        \end{figure*}

\begin{figure*}
                \centering
                \includegraphics[width=\hsize]{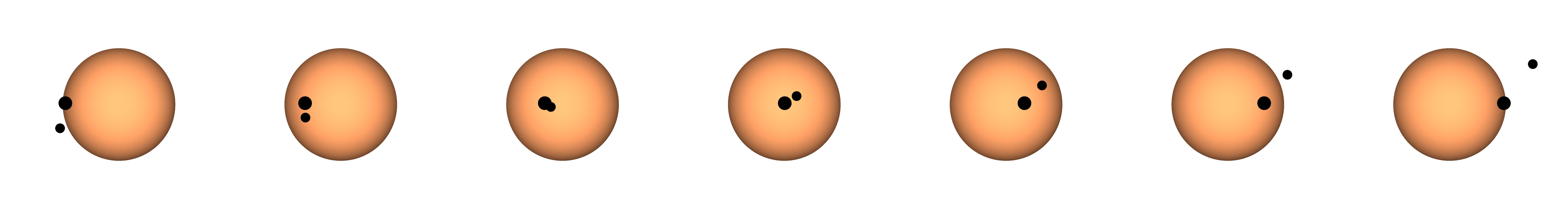}
                \caption{A series of snapshots showing a transit with the best-fit parameters from figure \ref{fig:kep51_fit}.} 
                \label{fig:kep51_snaps}
        \end{figure*}

        We also demonstrate the use of our model to estimate the posterior distribution for the model given the Kepler observations with MCMC. We use \texttt{emcee} \cite{ForemanMackey2019} to run 1,000 walkers for 44,000 steps. Figure \ref{fig:kep51_samples} shows a random selection of 100 light curves sampled from the posterior distribution of the model given the flux (here shown as blue points). Figure \ref{fig:kep51_posterior} shows the posterior distribution for the mutual inclination marginalized over all other parameters (right) alongside the joint distribution of the impact parameters for each planet. Our results are consistent with M14. In particular we infer a mutual inclination of $-33.0^{+21.5}_{-29.5}$ deg compared to the M14 value of $-25.3^{+6.2}_{-6.8}$ degrees. Because M14 do not describe the details of their MCMC algorithm it is difficult to explain the discrepancy between our relatively large error bars and their small error bars. In particular, M14 does not discuss what priors (if any) were used in their analysis. It should be noted however that in contrast to M14 our posterior admits models with low (though not zero) mutual inclination.
        
        Like M14 we find a low likelihood for the large impact parameters inferred from the phase-folded light curves of planets b and d. Figure \ref{fig:kep51_posterior} shows the joint posterior distribution for the impact parameters with the line $b_{K51b}=b_{K51d}$ and the values determined by the Kepler team. Since 2014, the Kepler-51 system has been observed by \textit{HST} which has yielded additional estimates of the impact parameters for Kepler-51b \& d \citep{Libby-Roberts2020}. We include these values in the figure as well, and note that these measurements are consistent with the impact parameters found by M14, and that the mean estimates of the impact parameters place both models in the low-probability region of the posterior. Figure \ref{fig:kep51_posterior} can be compared to Figure 4 in M14. The multiple modes seen in Figure \ref{fig:kep51_posterior} present an issue for the default \texttt{emcee} sampler that we used for this analysis as the chains have a low probability of crossing the low-likelihood valleys. As a result the chains tend to get stuck in the local high probability regions and do not fully mix between the modes. Given this, caution should be taken when interpreting our posteriors. In particular the apparent relative densities of each mode should not be taken to indicate a preference for one mode over another. We might partially rectify this limitation of our analysis by using Umbrella Sampling \citep{Torrie1977, Gilbert2022} to efficiently sample the low-likelihood regions between modes, or by employing an adaptive MCMC method designed for multi-modal likelihoods such as the Jumping Adaptive Multimodal Sampler \citep{Pompe2018}. Such an analysis is beyond the scope of this paper but may be the subject of future work. 
        
\section{Conclusions}
    \label{sec:conclusions}
    
    We have presented a solution for the flux during a mutual transit event in which two bodies overlap during their simultaneous transit of a star. It is mathematically equivalent to the solution given by \cite{pal2012}, but is parameterized differently and includes derivatives with respect to the positions and radii of both bodies. While \cite{pal2012} gives a general method for computing the flux blocked by $N$ overlapping bodies, our solution only applies for exactly two bodies. As a result we are able to give an explicit analytic solution for each of 16 different possible geometric cases. 
    
    We couple our solution to a dynamical model to build a full photodynamical model in a code which we call \texttt{gefera}. We give the user an option to model either a hierarchical system such as an exomoon/exoplanet system or hierarchical triple-star system or a confocal system such as two exoplanets orbiting the same star. We compare the speed of our implementation to \texttt{photodynam} and find that it performs several times faster even with the inclusion of the derivative computation. We also discuss the differences between our photodynamical model and \texttt{LUNA}, but because \texttt{LUNA} computes only an approximate solution and has not been made publicly available, a direct comparison is difficult to make. 
    
    As a demonstration of our method we conduct a brief re-analysis of a simultaneous transit of Kepler-51 b \& d, which was originally studied by \cite{Masuda2014}. We come to the conclusion that \cite{Masuda2014} is correct in assessing that an overlapping planet-planet transit is an unlikely explanation of the data, we further agree with their conclusion that an overlapping transit explanation is inconsistent with the measured impact parameters of the two planets. 
    
    The code \texttt{gefera} is made publicly available as a pip-installable python package. It is our hope that the release of this package will lower barriers to conducting exomoon searches, carrying out independent analyses of putative exomoon transits and planet-planet mutual transits, and modeling transiting triple-star or circumbinary systems. 
    
\acknowledgements

This paper includes data collected by the Kepler mission and obtained from the MAST data archive at the Space Telescope Science Institute (STScI). Funding for the Kepler mission is provided by the NASA Science Mission Directorate. STScI is operated by the Association of Universities for Research in Astronomy, Inc., under NASA contract NAS 5–26555.

This research has made use of the NASA Exoplanet Archive, which is operated by the California Institute of Technology, under contract with the National Aeronautics and Space Administration under the Exoplanet Exploration Program.

The authors would like to thank René Heller and Michael Hippke for helpful comments on the manuscript and the referee for a thorough report.

We acknowledge support from NSF grant AST-1907342, NASA NExSS grant No.\ 80NSSC18K0829, and NASA XRP grant 80NSSC21K1111.

\software{Python, IPython \citep{ipython}, 
NumPy \citep{numpy}, 
Matplotlib \citep{matplotlib}, 
SciPy \citep{scipy}, 
Astropy \citep{astropy},
Lightkurve \citep{lightkurve},
Emcee \citep{ForemanMackey2019}
}

\facilities{Exoplanet Archive, Kepler}

\appendix  
    \section{Derivative of the flux with respect to $\phi$ in the limit of small center-of-star-moon separations}
    \label{appendix:small_bm}
    To see how the two factors of $b_m$ cancel out to resolve the apparent singularity in the limit $b_m\to0$, let us begin by identifying the factors of $b_m$ that appear in the series expansion of the numerator. The first of these comes from the factor of $\sin\theta$ in the numerator of Equations \eqref{eqn:dthetapm_dbp} and \eqref{eqn:dthetapm_dtheta} for which the first term of the Taylor series expansion is $b_m/b_p$. For Equation \eqref{eqn:dthetapm_dbpm} we first note that in the limit $b_m\to 0$ we have $b_{pm}\to b_p$, which allows us to factor out $b_p$ from the term $b_p\cos\theta-b_{pm}$ leaving a factor of $\cos\theta - 1$ in the numerator. The first term of the series expansion for $\cos\theta-1$ is $b_m^2/(2b_p^2)$, which fully cancels out the factors of $b_m$ in the denominator. 
            
    While this takes care of the singularity in Equation \eqref{eqn:dthetapm_dbpm}, we still need to find another factor of $b_m$ in the numerator of Equations \eqref{eqn:dthetapm_dbp} and \eqref{eqn:dthetapm_dtheta}. To see where this second factor comes from we look to the Taylor series expansion of the $\phi$-derivatives of each of the primitive integrals (Equations \eqref{eqn:prim_uniform}, \eqref{eqn:prim_linear}, \eqref{eqn:prim_quad}, and \eqref{eqn:prim_poly}). Because each integral is evaluated from $0$ to $\phi$ taking the derivative with respect to $\phi$ is simple -- by the fundamental theorem of calculus we recover the integrand itself evaluated at $\phi'=\phi$. The first term in the Taylor series expansion for small $b$ can be found by setting $b=0$. We immediately note that for each of these integrands setting $b=0$ eliminates all factors of $\phi$ and trigonometric functions thereof from the expressions. In other words we can define 
    \begin{equation}
    \label{eqn:taylor_bm}
        H_{m, n}(r) = \lim_{b_m\to0}\frac{\partial G_n(\phi, r_m, b_m)}{\partial \phi}
    \end{equation}
    to be the constant term of the Taylor series where $H_{m,n}$ is a function only of $r$. The integral along an arc of the moon intersected by the planet is computed from two evaluations of the primitive integrals as
    \begin{equation}
        \sum_{n=0}^N u_n\left[G_n(\phi_m^-, r_m, b_m)-G_n(\phi_m^+, r_m, b_m)\right].
    \end{equation} This term appears in cases E, F, H, and K of Table \ref{tbl:configurations}, which represent all the cases that can arise when $b_m=0$. Using the definition from Equation \eqref{eqn:taylor_bm} we can write the first term of the Taylor expansion for the derivative of this expression with respect to $\phi$ as 
    \begin{equation}
        \sum_{n=0}^N u_n\left[H_{m,n}(r_m)-H_{m,n}(r_m)\right] = 0.
    \end{equation} 
    We now see that the constant terms cancel out when the integral is evaluated along any arc of the moon in the limit that $b_m\to0$. All remaining terms of the Taylor expansion have at least one factor of $b_m$ in their numerators which is sufficient to cancel the factor of $b_m$ in the denominator of Equations \eqref{eqn:dthetapm_dbp} through \eqref{eqn:dthetapm_dtheta} when the chain rule is applied. 
            
    We can apply this same reasoning to the arc of the planet. In this case the only variable in the primitive integrals which depends on $b_m$ is the integration limit, $\phi$, since the $b$ in the integrands refers to $b_p$. In this case taking the limit as $b_m\to0$ is equivalent to taking the simultaneous limit as $\theta\to0$ and $b_{pm}\to b_p$. As we did for arcs along the moon we define 
    \begin{equation}
        \label{eqn:Taylor_bp}
        H_{p,n}(\phi, r_p, b_p) = \lim_{\theta-\to 0, b_{pm}\to b_p}\frac{\partial G_n(\phi, r_p, b_p)}{\partial\phi}
    \end{equation}
    Note that this expression is a function of $\phi$, unlike the corresponding limit for arcs along the moon in Equation \eqref{eqn:taylor_bm}. When $\theta\to0$ the integration limits along the arc become $\phi=\pm\phi_p$ (see Equation \eqref{eqn:phip}). We again consider cases E, F, H, and K as the relevant cases for $b_m=0$. Each of these cases includes a term 
    \begin{equation}
        \sum_{n=0}^N u_n\left[G_n(\phi_p^-, r_p, b_p)+G_n(\phi_p^+, r_p, b_p)\right].
    \end{equation} In some cases the expressions in Table \ref{tbl:configurations} must be rearranged to reveal this term, but it does arise in each case. Taking the limit of the $\phi$-derivative of this term and using the definition of $H_{p, n}$ from \eqref{eqn:Taylor_bp} we have
    \begin{eqnarray}
        \nonumber &&\sum_{n=0}^N u_n\left[H_{n,p}(-\phi_p, r_p, b_p) + H_{n,p}(\phi_p, r_p, b_p)\right] \\ \nonumber &=& \sum_{n=0}^N u_n \left[-H_{n,p}(\phi_p, r_p, b_p) + H_{n,p}(\phi_p, r_p, b_p)\right] \\ &=& 0
    \end{eqnarray}
    where we have used the fact that the derivative of an even function (such as $G_n$ which is even with respect to $\phi$) is an odd function to make the substitution $H_{n,p}(-\phi_p, r_p, b_p) = -H_{n,p}(\phi_p, r_p, b_p)$. Again we see that the constant terms of the expansion are eliminated when the integral is taken along a complete arc, and the remaining terms contain at least one factor of $b_m$ to cancel the factor of $b_m$ in the denominator of Equations \eqref{eqn:dthetapm_dbp} through \eqref{eqn:dthetapm_dtheta}.
            
    None of the remaining terms in cases E, F, H, or K depend on $\theta_m$, and therefore are not subject to the singularity in the derivatives of $\theta_m$. Therefore we do not need to account for factors of $b_m$ in any of these terms.
            
    Unfortunately some numerical error is introduced into the calculation by these subtractions and cancellations. In our testing we have found that this error is limited to the region in which $b_m$ is less than $10^{-6}$ stellar radii and is generally smaller than $10^{-5}$. It occurs only in the derivatives of parameters that control the position of the moon. 
    
    \section{Correction to Equations 37 in Kipping (2011) and 16 in Fewell (2006)}
    \label{appendix:fewell_correction}
            
\begin{figure}
                \centering
                \includegraphics[width=\hsize]{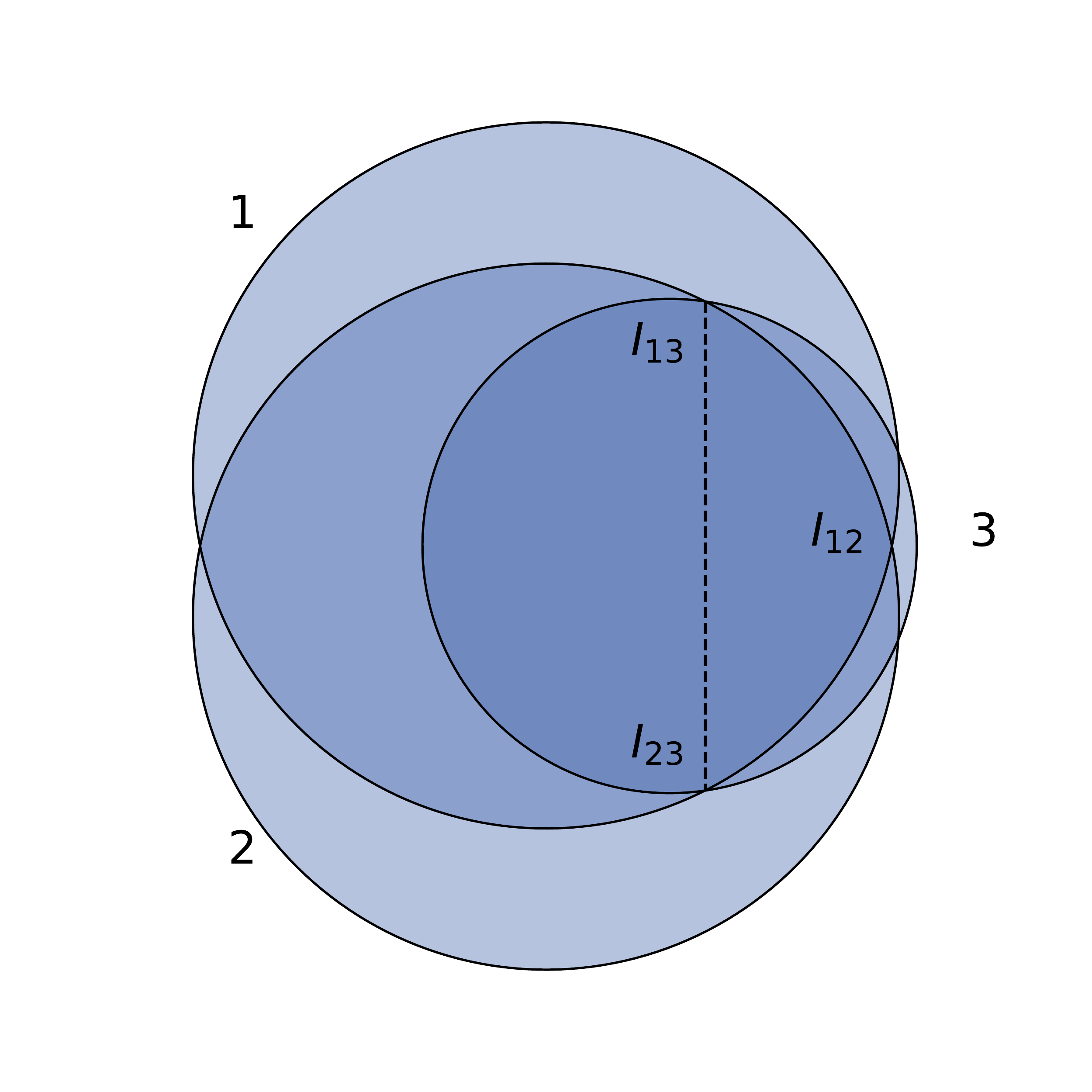}
                \caption{Three overlapping circles meeting the condition that more than half of the area of circle 3 is to the left of the chord connecting $I_{12}$ and $I_{13}$.} 
                \label{fig:fewell_overlap}
            \end{figure}

            This is a correction to the formula for the area of overlap of three circles when the chord connecting the intersections of the smallest circle with each of the larger circles spans more than half of the smallest circle. This configuration is shown in Figure \ref{fig:fewell_overlap} where the three circles are labeled $1$ through $3$ with circles $1$ and $2$ having the same radius and circle $3$ having a smaller radius. The intersections demarcating the area of overlap are labeled $I_{12}$, $I_{13}$ and $I_{23}$ with the subscript indicating the pair of circles that are intersecting. In Figure \ref{fig:fewell_overlap} the region to the left of the vertical dashed line contains more than half of circle $3$ (i.e. Equation (34) in \citealt{Kipping2011} is not true).
            
            \cite{Fewell2006} (and therefore \citealt{Kipping2011} as well) gives the area of the full region as
            \begin{eqnarray}
                \label{eqn:fewell}
                A_\mathrm{fewell} = \frac{\Delta}{4} + \sum_{i=1}^{3}r_i^2\mathrm{arcsin}\frac{c_i}{2r_i} - \frac{c_1}{4}\sqrt{4r_1^2-c_1^2} - \frac{c_2}{4}\sqrt{4r_2^2-c_2^2} + \frac{c_3}{4}\sqrt{4r_3^2-c_3^2}
            \end{eqnarray}
            where $\Delta$ is given by
            \begin{equation}
                \Delta = \sqrt{(c_1+c_2+c_3)(c_1+c_2-c_3)(c_1-c_2+c_3)(c_2+c_3-c_1)}
            \end{equation}
            and $\Delta/4$ is Heron's formula for the area of a triangle with side lengths $c_1$, $c_2$, and $c_3$.
            
\begin{figure*}
                \centering
                \includegraphics[width=\textwidth]{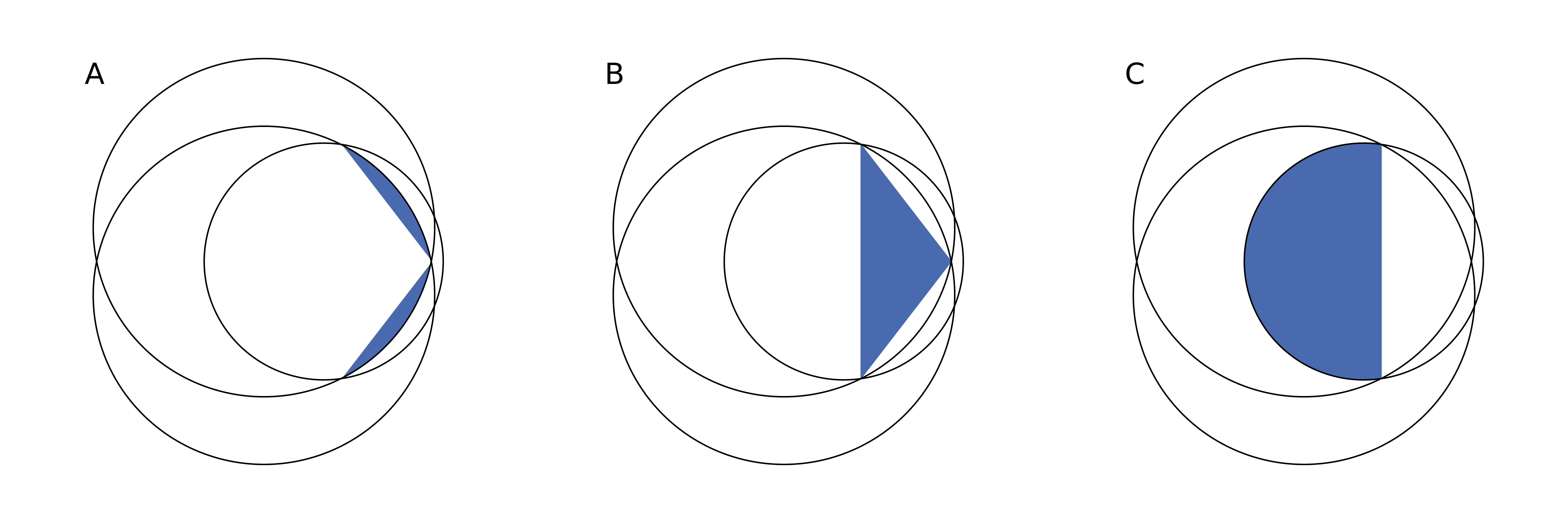}
                \caption{Subdivision of the region of overlap between three circles. The total area of overlap is found by adding the areas of each of these sub-regions. The expressions for each area are given in the text.} 
                \label{fig:fewell_correction}
            \end{figure*}

            The above formula is incorrect. The correct expression can be arrived at by breaking down the region of overlap into three sub-regions and considering the area of each separately. Figure \ref{fig:fewell_correction} highlights each of these regions. The first region, labeled A in Figure \ref{fig:fewell_correction} is formed by the two minor segments of circles 1 and 2 formed by the chords connecting $I_{12}$ to $I_{23}$ and $I_{13}$ to $I_{23}$. The general formula for the area of a minor segment is $A_\mathrm{segment} = r^2(\theta - \sin\theta)$ where $\theta$ is the angular length of the chord, which in this case yields the formula
            \begin{equation}
                A = \sum_{i=1}^{2}\left(r_i^2\mathrm{arcsin}\frac{c_i}{2r_i} - \frac{c_i}{4}\sqrt{4r_i^2-c_i^2}\right),
            \end{equation}
            where $r_i$ is the radius of circle $i$ and $c_i$ is the length of the chord segmenting circle $i$. This formula agrees with \cite{Fewell2006}, as does the formula for the area of the next region, region B in Figure \ref{fig:fewell_correction} which is the triangle formed by connecting the three points of intersection between the three circles. The area of this triangle is given by Heron's formula:
            \begin{equation}
                B = \frac{\Delta}{4}
            \end{equation}
            where $\Delta$ is defined as in Equation \eqref{eqn:fewell}
            
            The disagreement between our formula and \cite{Fewell2006} arises for the third sub-region, labeled region $C$ in \ref{fig:fewell_correction}, which is the remainder of the area of circle 3 after the minor segment formed by the chord connecting the intersections with circles 1 and 2 is subtracted. This can be expressed as
            \begin{eqnarray}
                C &=& \pi r_3^2 - (r_3^2\mathrm{arcsin}\frac{c_3}{2r_3} -\frac{c_3}{4}\sqrt{4r_3^2-c_3^2}) \\ \nonumber
                &=& r_3^2\left(\pi-\mathrm{arcsin}\frac{c_3}{2r_3}\right)+\frac{c_3}{4}\sqrt{4r_3^2-c_3^2}.
            \end{eqnarray}
            This differs from \cite{Fewell2006} and \cite{Kipping2011} in the inclusion of the $\pi r_3^2$ term as well as in the sign of the arcsin term. The corrected total area is then 
            \begin{eqnarray}
                \nonumber A_\mathrm{total} &=& \frac{\Delta}{4} + \sum_{i=1}^{2}\left(r_i^2\mathrm{arcsin}\frac{c_i}{2r_i} - \frac{c_i}{4}\sqrt{4r_i^2-c_i^2}\right) \\ \nonumber &+& r_3^2\left(\pi-\mathrm{arcsin}\frac{c_3}{2r_3}\right)+\frac{c_3}{4}\sqrt{4r_3^2-c_3^2}.
            \end{eqnarray}
            
\bibliography{main}

\begin{thebibliography}{}
\expandafter\ifx\csname natexlab\endcsname\relax\def\natexlab#1{#1}\fi
\providecommand{\url}[1]{\href{#1}{#1}}
\providecommand{\dodoi}[1]{doi:~\href{http://doi.org/#1}{\nolinkurl{#1}}}
\providecommand{\doeprint}[1]{\href{http://ascl.net/#1}{\nolinkurl{http://ascl.net/#1}}}
\providecommand{\doarXiv}[1]{\href{https://arxiv.org/abs/#1}{\nolinkurl{https://arxiv.org/abs/#1}}}

\bibitem[{{Agol} {et~al.}(2020){Agol}, {Luger}, \& {Foreman-Mackey}}]{Agol2020}
{Agol}, E., {Luger}, R., \& {Foreman-Mackey}, D. 2020, \aj, 159, 123,
  \dodoi{10.3847/1538-3881/ab4fee}

\bibitem[{{Astropy Collaboration} {et~al.}(2013){Astropy Collaboration},
  {Robitaille}, {Tollerud}, {Greenfield}, {Droettboom}, {Bray}, {Aldcroft},
  {Davis}, {Ginsburg}, {Price-Whelan}, {Kerzendorf}, {Conley}, {Crighton},
  {Barbary}, {Muna}, {Ferguson}, {Grollier}, {Parikh}, {Nair}, {Unther},
  {Deil}, {Woillez}, {Conseil}, {Kramer}, {Turner}, {Singer}, {Fox}, {Weaver},
  {Zabalza}, {Edwards}, {Azalee Bostroem}, {Burke}, {Casey}, {Crawford},
  {Dencheva}, {Ely}, {Jenness}, {Labrie}, {Lim}, {Pierfederici}, {Pontzen},
  {Ptak}, {Refsdal}, {Servillat}, \& {Streicher}}]{astropy}
{Astropy Collaboration}, {Robitaille}, T.~P., {Tollerud}, E.~J., {et~al.} 2013,
  \aap, 558, A33

\bibitem[{{Borkovits} {et~al.}(2013){Borkovits}, {Derekas}, {Kiss},
  {Kir{\'a}ly}, {Forg{\'a}cs-Dajka}, {B{\'\i}r{\'o}}, {Bedding}, {Bryson},
  {Huber}, \& {Szab{\'o}}}]{Borkovits2013}
{Borkovits}, T., {Derekas}, A., {Kiss}, L.~L., {et~al.} 2013, \mnras, 428,
  1656, \dodoi{10.1093/mnras/sts146}

\bibitem[{{Borkovits} {et~al.}(2019){Borkovits}, {Rappaport}, {Kaye},
  {Isaacson}, {Vanderburg}, {Howard}, {Kristiansen}, {Omohundro},
  {Schwengeler}, {Terentev}, {Shporer}, {Relles}, {Villanueva}, {Tan},
  {Col{\'o}n}, {Blex}, {Haas}, {Cochran}, \& {Endl}}]{Borkovits2019a}
{Borkovits}, T., {Rappaport}, S., {Kaye}, T., {et~al.} 2019, \mnras, 483, 1934,
  \dodoi{10.1093/mnras/sty3157}

\bibitem[{{Borkovits} {et~al.}(2020){Borkovits}, {Rappaport}, {Tan},
  {Gagliano}, {Jacobs}, {Huang}, {Mitnyan}, {Hambsch}, {Kaye}, {Maxted},
  {P{\'a}l}, \& {Schmitt}}]{Borkovits2020}
{Borkovits}, T., {Rappaport}, S.~A., {Tan}, T.~G., {et~al.} 2020, \mnras, 496,
  4624, \dodoi{10.1093/mnras/staa1817}

\bibitem[{{Borkovits} {et~al.}(2022){Borkovits}, {Mitnyan}, {Rappaport},
  {Pribulla}, {Powell}, {Kostov}, {B{\'\i}r{\'o}}, {Cs{\'a}nyi}, {Garai},
  {Gary}, {Kaye}, {Kom{\v{z}}{\'\i}k}, {Terentev}, {Omohundro}, {Gagliano},
  {Jacobs}, {Kristiansen}, {LaCourse}, {Schwengeler}, {Czavalinga}, {Seli},
  {Huang}, {P{\'a}l}, {Vanderburg}, {Rodriguez}, \& {Stevens}}]{Borkovits2022}
{Borkovits}, T., {Mitnyan}, T., {Rappaport}, S.~A., {et~al.} 2022, \mnras, 510,
  1352, \dodoi{10.1093/mnras/stab3397}

\bibitem[{{Brakensiek} \& {Ragozzine}(2016)}]{Brakensiek2016}
{Brakensiek}, J., \& {Ragozzine}, D. 2016, \apj, 821, 47,
  \dodoi{10.3847/0004-637X/821/1/47}

\bibitem[{Bulirsch(1965)}]{Bulirsch1965el1el2}
Bulirsch, R. 1965, Numerische Mathematik, 7, 78

\bibitem[{Bulirsch(1969{\natexlab{a}})}]{Bulirsch1969el3}
---. 1969{\natexlab{a}}, Numerische Mathematik, 13, 305

\bibitem[{Bulirsch(1969{\natexlab{b}})}]{Bulirsch1969lincombo}
---. 1969{\natexlab{b}}, Numerische Mathematik, 13, 266

\bibitem[{{Byrd} \& {Friedman}(1954)}]{Byrd1954}
{Byrd}, P.~F., \& {Friedman}, M. 1954, Mitteilungen der Astronomischen
  Gesellschaft Hamburg, 5, 99

\bibitem[{{Carter} {et~al.}(2008){Carter}, {Yee}, {Eastman}, {Gaudi}, \&
  {Winn}}]{Carter2008}
{Carter}, J.~A., {Yee}, J.~C., {Eastman}, J., {Gaudi}, B.~S., \& {Winn}, J.~N.
  2008, \apj, 689, 499, \dodoi{10.1086/592321}

\bibitem[{{Carter} {et~al.}(2011){Carter}, {Fabrycky}, {Ragozzine}, {Holman},
  {Quinn}, {Latham}, {Buchhave}, {Van Cleve}, {Cochran}, {Cote}, {Endl},
  {Ford}, {Haas}, {Jenkins}, {Koch}, {Li}, {Lissauer}, {MacQueen}, {Middour},
  {Orosz}, {Rowe}, {Steffen}, \& {Welsh}}]{Carter2011}
{Carter}, J.~A., {Fabrycky}, D.~C., {Ragozzine}, D., {et~al.} 2011, Science,
  331, 562, \dodoi{10.1126/science.1201274}

\bibitem[{{Carter} {et~al.}(2012){Carter}, {Agol}, {Chaplin}, {Basu},
  {Bedding}, {Buchhave}, {Christensen-Dalsgaard}, {Deck}, {Elsworth},
  {Fabrycky}, {Ford}, {Fortney}, {Hale}, {Handberg}, {Hekker}, {Holman},
  {Huber}, {Karoff}, {Kawaler}, {Kjeldsen}, {Lissauer}, {Lopez}, {Lund},
  {Lundkvist}, {Metcalfe}, {Miglio}, {Rogers}, {Stello}, {Borucki}, {Bryson},
  {Christiansen}, {Cochran}, {Geary}, {Gilliland}, {Haas}, {Hall}, {Howard},
  {Jenkins}, {Klaus}, {Koch}, {Latham}, {MacQueen}, {Sasselov}, {Steffen},
  {Twicken}, \& {Winn}}]{Carter2012}
{Carter}, J.~A., {Agol}, E., {Chaplin}, W.~J., {et~al.} 2012, Science, 337,
  556, \dodoi{10.1126/science.1223269}

\bibitem[{{Doyle} {et~al.}(2011){Doyle}, {Carter}, {Fabrycky}, {Slawson},
  {Howell}, {Winn}, {Orosz}, {P{\v{r}}sa}, {Welsh}, {Quinn}, {Latham},
  {Torres}, {Buchhave}, {Marcy}, {Fortney}, {Shporer}, {Ford}, {Lissauer},
  {Ragozzine}, {Rucker}, {Batalha}, {Jenkins}, {Borucki}, {Koch}, {Middour},
  {Hall}, {McCauliff}, {Fanelli}, {Quintana}, {Holman}, {Caldwell}, {Still},
  {Stefanik}, {Brown}, {Esquerdo}, {Tang}, {Furesz}, {Geary}, {Berlind},
  {Calkins}, {Short}, {Steffen}, {Sasselov}, {Dunham}, {Cochran}, {Boss},
  {Haas}, {Buzasi}, \& {Fischer}}]{Doyle2011}
{Doyle}, L.~R., {Carter}, J.~A., {Fabrycky}, D.~C., {et~al.} 2011, Science,
  333, 1602, \dodoi{10.1126/science.1210923}

\bibitem[{Fewell(2006)}]{Fewell2006}
Fewell, M.~P. 2006, Area of common overlap of three circles, Tech. rep.,
  Defence Science and Technology Group

\bibitem[{{Foreman-Mackey} {et~al.}(2019){Foreman-Mackey}, {Farr}, {Sinha},
  {Archibald}, {Hogg}, {Sanders}, {Zuntz}, {Williams}, {Nelson}, {de
  Val-Borro}, {Erhardt}, {Pashchenko}, \& {Pla}}]{ForemanMackey2019}
{Foreman-Mackey}, D., {Farr}, W., {Sinha}, M., {et~al.} 2019, The Journal of
  Open Source Software, 4, 1864, \dodoi{10.21105/joss.01864}

\bibitem[{{Foreman-Mackey} {et~al.}(2021){Foreman-Mackey}, {Luger}, {Agol},
  {Barclay}, {Bouma}, {Brandt}, {Czekala}, {David}, {Dong}, {Gilbert},
  {Gordon}, {Hedges}, {Hey}, {Morris}, {Price-Whelan}, \&
  {Savel}}]{ForemanMackey2021}
{Foreman-Mackey}, D., {Luger}, R., {Agol}, E., {et~al.} 2021, The Journal of
  Open Source Software, 6, 3285, \dodoi{10.21105/joss.03285}

\bibitem[{{Gilbert}(2022)}]{Gilbert2022}
{Gilbert}, G.~J. 2022, \aj, 163, 111, \dodoi{10.3847/1538-3881/ac45f4}

\bibitem[{Gordon(2022)}]{gefera_zenodo}
Gordon, T. 2022, tagordon/gefera: Gefera 0.1, v0.1,  Zenodo,
  \dodoi{10.5281/zenodo.6817200}.
\newblock \url{https://doi.org/10.5281/zenodo.6817200}

\bibitem[{{Hippke} \& {Heller}(2022)}]{Hippke2022}
{Hippke}, M., \& {Heller}, R. 2022, \aap, 662, A37,
  \dodoi{10.1051/0004-6361/202243129}

\bibitem[{{Hirano} {et~al.}(2012){Hirano}, {Narita}, {Sato}, {Takahashi},
  {Masuda}, {Takeda}, {Aoki}, {Tamura}, \& {Suto}}]{Hirano2012}
{Hirano}, T., {Narita}, N., {Sato}, B., {et~al.} 2012, \apjl, 759, L36,
  \dodoi{10.1088/2041-8205/759/2/L36}

\bibitem[{Hoffman \& Gelman(2014)}]{Hoffman2014}
Hoffman, M.~D., \& Gelman, A. 2014, Journal of Machine Learning Research, 15,
  1593

\bibitem[{Hunter(2007)}]{matplotlib}
Hunter, J.~D. 2007, Computing In Science \& Engineering, 9, 90

\bibitem[{{Kahan}(2000)}]{kahan2000}
{Kahan}, W. 2000, Tech. rep., University of California, Berkeley

\bibitem[{{Kipping} {et~al.}(2022){Kipping}, {Bryson}, {Burke}, {Christiansen},
  {Hardegree-Ullman}, {Quarles}, {Hansen}, {Szul{\'a}gyi}, \&
  {Teachey}}]{Kipping2022}
{Kipping}, D., {Bryson}, S., {Burke}, C., {et~al.} 2022, Nature Astronomy,
  \dodoi{10.1038/s41550-021-01539-1}

\bibitem[{{Kipping}(2010)}]{Kippin2010}
{Kipping}, D.~M. 2010, \mnras, 408, 1758,
  \dodoi{10.1111/j.1365-2966.2010.17242.x}

\bibitem[{{Kipping}(2011)}]{Kipping2011}
---. 2011, \mnras, 416, 689, \dodoi{10.1111/j.1365-2966.2011.19086.x}

\bibitem[{{Libby-Roberts} {et~al.}(2020){Libby-Roberts}, {Berta-Thompson},
  {D{\'e}sert}, {Masuda}, {Morley}, {Lopez}, {Deck}, {Fabrycky}, {Fortney},
  {Line}, {Sanchis-Ojeda}, \& {Winn}}]{Libby-Roberts2020}
{Libby-Roberts}, J.~E., {Berta-Thompson}, Z.~K., {D{\'e}sert}, J.-M., {et~al.}
  2020, \aj, 159, 57, \dodoi{10.3847/1538-3881/ab5d36}

\bibitem[{{Lightkurve Collaboration} {et~al.}(2018){Lightkurve Collaboration},
  {Cardoso}, {Hedges}, {Gully-Santiago}, {Saunders}, {Cody}, {Barclay}, {Hall},
  {Sagear}, {Turtelboom}, {Zhang}, {Tzanidakis}, {Mighell}, {Coughlin}, {Bell},
  {Berta-Thompson}, {Williams}, {Dotson}, \& {Barentsen}}]{lightkurve}
{Lightkurve Collaboration}, {Cardoso}, J.~V.~d.~M., {Hedges}, C., {et~al.}
  2018, {Lightkurve: Kepler and TESS time series analysis in Python},
  Astrophysics Source Code Library.
\newblock \doeprint{1812.013}

\bibitem[{{Luger} {et~al.}(2019){Luger}, {Agol}, {Foreman-Mackey}, {Fleming},
  {Lustig-Yaeger}, \& {Deitrick}}]{Luger2019}
{Luger}, R., {Agol}, E., {Foreman-Mackey}, D., {et~al.} 2019, \aj, 157, 64,
  \dodoi{10.3847/1538-3881/aae8e5}

\bibitem[{{Luger} {et~al.}(2017){Luger}, {Lustig-Yaeger}, \&
  {Agol}}]{Luger2017}
{Luger}, R., {Lustig-Yaeger}, J., \& {Agol}, E. 2017, \apj, 851, 94,
  \dodoi{10.3847/1538-4357/aa9c43}

\bibitem[{{Mandel} \& {Agol}(2002)}]{Mandel2002}
{Mandel}, K., \& {Agol}, E. 2002, \apjl, 580, L171, \dodoi{10.1086/345520}

\bibitem[{{Martin} {et~al.}(2019){Martin}, {Fabrycky}, \&
  {Montet}}]{Martin2019}
{Martin}, D.~V., {Fabrycky}, D.~C., \& {Montet}, B.~T. 2019, \apjl, 875, L25,
  \dodoi{10.3847/2041-8213/ab0aea}

\bibitem[{{Masuda}(2014)}]{Masuda2014}
{Masuda}, K. 2014, \apj, 783, 53, \dodoi{10.1088/0004-637X/783/1/53}

\bibitem[{{Masuda} {et~al.}(2013){Masuda}, {Hirano}, {Taruya}, {Nagasawa}, \&
  {Suto}}]{Masuda2013}
{Masuda}, K., {Hirano}, T., {Taruya}, A., {Nagasawa}, M., \& {Suto}, Y. 2013,
  \apj, 778, 185, \dodoi{10.1088/0004-637X/778/2/185}

\bibitem[{{Mitnyan} {et~al.}(2020){Mitnyan}, {Borkovits}, {Rappaport},
  {P{\'a}l}, \& {Maxted}}]{Mitnyan2020}
{Mitnyan}, T., {Borkovits}, T., {Rappaport}, S.~A., {P{\'a}l}, A., \& {Maxted},
  P.~F.~L. 2020, \mnras, 498, 6034, \dodoi{10.1093/mnras/staa2762}

\bibitem[{{NASA Exoplanet Archive}(YYYY)}]{planetary_systems_table}
{NASA Exoplanet Archive}. YYYY, Planetary Systems, Version: YYYY-MM-DD HH:MM,
  NExScI-Caltech/IPAC, \dodoi{10.26133/NEA12}.
\newblock \url{https://catcopy.ipac.caltech.edu/dois/doi.php?id=10.26133/NEA12}

\bibitem[{Oliphant(2007)}]{numpy}
Oliphant, T.~E. 2007, Computing in Science Engineering, 9, 10

\bibitem[{{P{\'a}l}(2012)}]{pal2012}
{P{\'a}l}, A. 2012, \mnras, 420, 1630, \dodoi{10.1111/j.1365-2966.2011.20151.x}

\bibitem[{P\'erez \& Granger(2007)}]{ipython}
P\'erez, F., \& Granger, B.~E. 2007, Computing in Science and Engineering, 9,
  21

\bibitem[{{Pompe} {et~al.}(2018){Pompe}, {Holmes}, \&
  {{\L}atuszy{\'n}ski}}]{Pompe2018}
{Pompe}, E., {Holmes}, C., \& {{\L}atuszy{\'n}ski}, K. 2018, arXiv e-prints,
  arXiv:1812.02609.
\newblock \doarXiv{1812.02609}

\bibitem[{{Price} \& {Rogers}(2014)}]{Price2014}
{Price}, E.~M., \& {Rogers}, L.~A. 2014, \apj, 794, 92,
  \dodoi{10.1088/0004-637X/794/1/92}

\bibitem[{{Ragozzine} \& {Holman}(2010)}]{Ragozzine2010}
{Ragozzine}, D., \& {Holman}, M.~J. 2010, arXiv e-prints, arXiv:1006.3727.
\newblock \doarXiv{1006.3727}

\bibitem[{{Raposo-Pulido} \& {Pel{\'a}ez}(2017)}]{RaposoPulido2017}
{Raposo-Pulido}, V., \& {Pel{\'a}ez}, J. 2017, \mnras, 467, 1702,
  \dodoi{10.1093/mnras/stx138}

\bibitem[{Salvatier {et~al.}(2016)Salvatier, Wiecki, \&
  Fonnesbeck}]{Salvatier2016}
Salvatier, J., Wiecki, T.~V., \& Fonnesbeck, C. 2016, {PeerJ} Computer Science,
  2, e55, \dodoi{10.7717/peerj-cs.55}

\bibitem[{{Short} {et~al.}(2018){Short}, {Orosz}, {Windmiller}, \&
  {Welsh}}]{Short2018}
{Short}, D.~R., {Orosz}, J.~A., {Windmiller}, G., \& {Welsh}, W.~F. 2018, \aj,
  156, 297, \dodoi{10.3847/1538-3881/aae889}

\bibitem[{{Teachey} \& {Kipping}(2018)}]{Teachey2018}
{Teachey}, A., \& {Kipping}, D.~M. 2018, Science Advances, 4, eaav1784,
  \dodoi{10.1126/sciadv.aav1784}

\bibitem[{{Torrie} \& {Valleau}(1977)}]{Torrie1977}
{Torrie}, G.~M., \& {Valleau}, J.~P. 1977, Journal of Computational Physics,
  23, 187, \dodoi{10.1016/0021-9991(77)90121-8}

\bibitem[{{Vallisneri}(2008)}]{Vallisneri2008}
{Vallisneri}, M. 2008, \prd, 77, 042001, \dodoi{10.1103/PhysRevD.77.042001}

\bibitem[{{Virtanen} {et~al.}(2020){Virtanen}, {Gommers}, {Oliphant},
  {Haberland}, {Reddy}, {Cournapeau}, {Burovski}, {Peterson}, {Weckesser},
  {Bright}, {van der Walt}, {Brett}, {Wilson}, {Jarrod Millman}, {Mayorov},
  {Nelson}, {Jones}, {Kern}, {Larson}, {Carey}, {Polat}, {Feng}, {Moore}, {Vand
  erPlas}, {Laxalde}, {Perktold}, {Cimrman}, {Henriksen}, {Quintero}, {Harris},
  {Archibald}, {Ribeiro}, {Pedregosa}, {van Mulbregt}, \&
  {Contributors}}]{scipy}
{Virtanen}, P., {Gommers}, R., {Oliphant}, T.~E., {et~al.} 2020, Nature
  Methods, 17, 261, \dodoi{https://doi.org/10.1038/s41592-019-0686-2}

\bibitem[{{Welsh} {et~al.}(2012){Welsh}, {Orosz}, {Carter}, {Fabrycky}, {Ford},
  {Lissauer}, {Pr{\v{s}}a}, {Quinn}, {Ragozzine}, {Short}, {Torres}, {Winn},
  {Doyle}, {Barclay}, {Batalha}, {Bloemen}, {Brugamyer}, {Buchhave},
  {Caldwell}, {Caldwell}, {Christiansen}, {Ciardi}, {Cochran}, {Endl},
  {Fortney}, {Gautier}, {Gilliland}, {Haas}, {Hall}, {Holman}, {Howard},
  {Howell}, {Isaacson}, {Jenkins}, {Klaus}, {Latham}, {Li}, {Marcy}, {Mazeh},
  {Quintana}, {Robertson}, {Shporer}, {Steffen}, {Windmiller}, {Koch}, \&
  {Borucki}}]{Welsh2012}
{Welsh}, W.~F., {Orosz}, J.~A., {Carter}, J.~A., {et~al.} 2012, \nat, 481, 475,
  \dodoi{10.1038/nature10768}

\bibitem[{Willard {et~al.}(2021)Willard, Osthege, Ho, Vieira, Wiecki,
  Foreman-Mackey, Chaudhari, Legrand, Kumar, Lao, Abril-Pla, Fonnesbeck,
  Goldman, \& Gorelli}]{Willard2021}
Willard, B.~T., Osthege, M., Ho, G., {et~al.} 2021, pymc-devs/aesara:,
  rel-2.0.7,  Zenodo, \dodoi{10.5281/zenodo.4695331}.
\newblock \url{https://doi.org/10.5281/zenodo.4695331}

\bibitem[{{Wilson} \& {Devinney}(1971)}]{Wilson1971}
{Wilson}, R.~E., \& {Devinney}, E.~J. 1971, \apj, 166, 605,
  \dodoi{10.1086/150986}

\bibitem[{{Winn}(2010)}]{Winn2010}
{Winn}, J.~N. 2010, in Exoplanets, ed. S.~{Seager}, 55--77

\end{thebibliography}



\end{document}